\documentclass[sigconf, nonacm]{acmart}

\usepackage{graphicx}
\usepackage{balance} 
\usepackage{todonotes}
\usepackage[multiple]{footmisc}
\usepackage{url}
\usepackage{textcomp}
\usepackage{listings}
\usepackage{caption}
\usepackage{subcaption}
\usepackage[group-separator={,},group-minimum-digits={3}]{siunitx}
\usepackage{enumitem}
\usepackage{amsmath,bm}
\usepackage{eurosym}
\usepackage{multicol}
\usepackage{multirow}

\AtBeginDocument{%
	}

\setcopyright{acmcopyright}
\copyrightyear{2018}
\acmYear{2018}
\acmDOI{XXXXXXX.XXXXXXX}

\acmConference[Conference acronym 'XX]{Make sure to enter the correct
	conference title from your rights confirmation emai}{June 03--05,
	2018}{Woodstock, NY}
\acmPrice{15.00}
\acmISBN{978-1-4503-XXXX-X/18/06}

\newcommand{\name}{$\text{Nextia}_{\text{JD}}$}

\begin{document}

\title{Measuring and Predicting the Quality of a Join for Data Discovery}

\author{Sergi Nadal, Raquel Panadero, Javier Flores, Oscar Romero}
\affiliation{%
	\institution{Universitat Politècnica de Catalunya, BarcelonaTech}
	\city{Barcelona}
	\country{Spain}
}
\email{{sergi.nadal| raquel.panadero | javier.flores | oscar.romero} @upc.edu}

\begin{abstract}
We study the problem of discovering joinable datasets at scale. We approach the problem from a learning perspective relying on profiles. These are succinct representations that capture the underlying characteristics of the schemata and data values of datasets, which can be efficiently extracted in a distributed and parallel fashion. Profiles are then compared, to predict the quality of a join operation among a pair of attributes from different datasets. In contrast to the state-of-the-art, we define a novel notion of join quality that relies on a metric considering both the containment and cardinality proportion between join candidate attributes. We implement our approach in a system called NextiaJD, and present experiments to show the predictive performance and computational efficiency of our method. Our experiments show that NextiaJD obtains greater predictive performance to that of hash-based methods while we are able to scale-up to larger volumes of data.
\end{abstract}

\maketitle

\begingroup\small\noindent\raggedright\textbf{Artifact Availability:} The source code, data, and/or other artifacts are available at \url{https://www.essi.upc.edu/dtim/NextiaJD/}.
	\endgroup

\section{Introduction} \label{sec:intro}

Data discovery is the broad process of navigating a large set of data sources in order to find relevant datasets and meaningful relationships among them \cite{DBLP:conf/icde/FernandezMQEIMO18,DBLP:conf/icde/BogatuFP020}. Discovery and integration of datasets is nowadays a largely manual and arduous task that consumes up to 80\% of a data scientists' time \cite{DBLP:journals/debu/StonebrakerI18}. This only gets aggravated by the proliferation of large repositories of heterogeneous data, such as \textit{data lakes} \cite{DBLP:journals/pvldb/NargesianZMPA19} or open data-related initiatives \cite{DBLP:journals/debu/MillerNZCPA18}. 
Due to the unprecedented large-scale volumes of heterogeneous data sources, manual data discovery becomes an unfeasible task that calls for automation \cite{DBLP:conf/pods/GolshanHMT17}. In this paper, we focus on the problem of discovering joinable attributes among datasets in a data lake. 

As an illustrative example of the challenges we face, take the reference dataset ($D_{ref}$) depicted in Table \ref{tab:happinessExample}. Assume we have a collection of other datasets available in the same data lake such as those depicted in Table \ref{tab:table_example_joins}. In such setting, we aim at finding joinable combinations of pairs of attributes from the reference dataset to all the rest. A first observation is that purely schema-based methods, such as LogMap \cite{DBLP:conf/semweb/Jimenez-RuizG11}, would fail to propose the combination $D_{ref}.Country = D_1.X$ due to the lack of embedded semantics in the schema of $D_1$. Thus, we must also take into account the data values. 
Note, however, that checking only data values might result in proposing the combination $D_{ref}.Schengen = D_2.Discount$, which is definitely not relevant for analysis. Furthermore, given positive pairs (i.e., likely to be meaningfully joinable), such as $D_{ref}.Country = D_1.X$ and $D_{ref}.Country = D_2.Country$, there should be a clear criteria to rank them (i.e., suggest which one is \emph{better}). Ranking is relevant in data lake scenarios, where independent data files must be crossed. Most of the times, in such scenarios, the cardinality of such files is not excessively large, but their order (i.e., number of columns / attributes) tend to be. Consequently, current approaches tend to propose too many joinable pairs of attributes, which is overwhelming for the end-user validating them. 

\begin{table}
	\centering
		\begin{tabular}{|c|c|c|}
			\hline
			\textbf{Country} & \textbf{Happiness score} & \textbf{Schengen} \\
			\hline
			Mexico                                    & 6.595           & N  \\
			\hline
			Spain                                     & 6.354           & Y  \\
			\hline
			United States                             & 6.892           & N  \\
			\hline
			France                                    & 6.592           & Y  \\
			\hline
		\end{tabular}
		\caption{\label{tab:happinessExample} $D_{ref}$ -- Happiness score per country in 2019}
\end{table}

The problem of finding joinable attributes among datasets is nowadays a topic of major interest for the data management community \cite{DBLP:conf/sigmod/BalazinskaCAFKS20,DBLP:journals/sigmod/AbadiAABBBBCCDD19}. We distinguish three approaches: \textit{comparison by value}, \textit{comparison by hash} and \textit{comparison by profile}. Table \ref{tab:relatedWork}, overviews recent contributions. Comparison by value relies on auxiliary data structures such as inverted indices or dictionaries to minimize the lookup cost.
Alternatively, comparison by hash expects that the signature of values under locality-sensitive hashing schemes will collision in the same bucket, also employing index structures for efficient threshold index. Comparison by profile methods leverage on profiles extracted from datasets and their attributes, which are used to predict whether a pair of attributes will join.

\begin{table}[htbp]
	\centering
		\begin{tabular}{|c|c|c|}
			\hline
			\multicolumn{3}{|c|}{\textbf{Search accuracy}} \\
			\multicolumn{3}{|c|}{{ Exact \dotfill Approximate  \hfill}}                                                                                                                                                                              \\ \hline
			\begin{tabular}[c]{@{}c@{}}Comp. by value\\ \cite{DBLP:journals/pvldb/DengKMS17,DBLP:conf/sigmod/ZhuDNM19, DBLP:journals/tods/XiaoWLYW11}\end{tabular} & \begin{tabular}[c]{@{}c@{}}Comp. by hash\\ \cite{DBLP:conf/sequences/Broder97,DBLP:conf/icde/FernandezMNM19,DBLP:conf/icde/YangZZH19,DBLP:journals/pvldb/ZhuNPM16,DBLP:conf/icde/FernandezAKYMS18,DBLP:conf/icde/BogatuFP020}\end{tabular} & \begin{tabular}[c]{@{}c@{}}Comp. by profile\\ \cite{DBLP:journals/debu/ChenGHTD18,DBLP:journals/ml/DoanDH03,DBLP:conf/esws/KejriwalM15a,DBLP:conf/icde/DongT0O21,DBLP:journals/pvldb/BharadwajGBG21,DBLP:journals/corr/abs-2212-14155}\end{tabular} \\ \hline
			\multicolumn{3}{|c|}{{Expensive\dotfill Efficient \hfill }}                                                                                                                                                          \\
			\multicolumn{3}{|c|}{\textbf{Algorithmic complexity}} \\ \hline
		\end{tabular}
		\captionof{table}{Overview of approaches by technique, arranged according to accuracy and algorithmic complexity} \label{tab:relatedWork}
\end{table}

\begin{table*}[t!]
	\begin{center}
		\begin{subtable}{.25\linewidth}
			\centering
			\caption{\label{tab:populationExample} $D_1$ -- Population per country}
			\resizebox{\textwidth}{!}{
				\begin{tabular}{|c|c|c|}
					\hline
					\textbf{X} & \textbf{Y} & \textbf{Z} \\
					\hline
					Spain & 47M & 2020 \\
					\hline
					United States & 330M & 2020 \\
					\hline
					Mexico & 123M & 2020 \\
					\hline
					Germany & 83M & 2020 \\
					\hline
				\end{tabular}
			}
		\end{subtable}
		\begin{subtable}{.01\linewidth}
		\end{subtable}
		\begin{subtable}{.35\linewidth}
			\caption{\label{tab:storesExample} $D_2$ -- Customer satisfaction per store location}
			\centering
			\resizebox{\textwidth}{!}{
				\begin{tabular}{|c|c|c|c|}
					\hline
					\textbf{Country} & \textbf{Location} & \textbf{Discount} & \textbf{ \begin{tabular}[c]{@{}>{}c@{}}Customer \\satisfaction \end{tabular} } \\
					\hline
					United States & New York & Y & 7.7 \\
					\hline
					United States & Chicago & N & 8.5 \\
					\hline
					United States & Seattle & N & 8  \\
					\hline
					United States & Houston & Y & 7.7 \\
					\hline
				\end{tabular}
			}
		\end{subtable}
		\begin{subtable}{.01\linewidth}
			\hspace*{\fill}
		\end{subtable}%
		\begin{subtable}{.375\linewidth}
			\centering
			\caption{\label{tab:expectancyExample} $D_3$ -- Average life expectancy}
			\resizebox{\textwidth}{!}{
				\begin{tabular}{|c|c|c|}
					\hline
					\textbf{Nation} & \textbf{ \begin{tabular}[c]{@{}>{}c@{}}Life expectancy \\(Women) \end{tabular} } & \textbf{ \begin{tabular}[c]{@{}>{}c@{}}Life expectancy \\(Men) \end{tabular} } \\
					\hline
					MX & 77.8 & 72.1 \\
					\hline
					SP & 86.1 & 86.1 \\
					\hline
					CA & 82.2 & 72.3 \\
					\hline
					US & 81.4 & 76.3 \\
					\hline
					BR & 79.4 & 72 \\
					\hline
				\end{tabular}
			}
		\end{subtable}
	\end{center}
	\caption{ \label{tab:table_example_joins} Three additional datasets of a toy-example repository ($D_1$, $D_2$ and $D_3$)}
\end{table*}

\subsection{Data discovery at scale}

Unfortunately, as we experimentally show in Section \ref{sec:experiments}, the state-of-the-art in data discovery does not meet the expectations for large-scale scenarios. Unlike traditional relational databases, these are characterized by
\textit{a)} a wide heterogeneity among datasets (e.g., large differences on the number of attributes and / or their cardinalities);
\textit{b)} massive number of datasets; and
\textit{c)} the presence of a variety of topics, or domains.
Overall, these distinguishing features, which we discuss as follows, deem current solutions ineffective due to their inability to scale-up and the low quality in rankings they provide. 

\medskip

\noindent\textbf{Inability to scale-up.} 
Solutions that yield exact results (i.e., comparison by value) quickly suffer from scalability problems. Indeed, assuming that the sets of values for a pair of attributes $A$, $B$ are maintained in memory as dictionaries, the complexity of computing their containment (i.e., the inclusion coefficient) is $O(min(|A|,|B|))$. 
Alternatively, those solutions that discover joinable pairs with a bounded error (i.e., comparison by hash) require the construction and maintenance of index structures for efficient lookup. This is a task that becomes highly demanding in terms of computing resources on large-scale datasets. In fact, as we have empirically observed and discuss in Section \ref{sec:experiments}, the available implementations on comparison by hash fail to handle datasets of few GBs overall. Another drawback of such approaches is that the estimation of similarity measures like containment is highly imprecise when the cardinality (i.e., the number of distinct values) of an attribute is comparatively larger than the other's \cite{DBLP:journals/pvldb/NaziDNC18}, which is a common characteristic in real-world large-scale applications. As a result, the precision of current approaches is highly affected due to the large number of false positives and worsened in large-scale settings.

\medskip

\noindent\textbf{Low quality in rankings.} 
Comparison by hash and profile solutions aim at predicting set-based measures such as the inclusion coefficient (i.e., containment), denoted $C$, or the Jaccard index, denoted $J$, using locality-sensitive hashing techniques such as MinHash \cite{DBLP:conf/sequences/Broder97} or random projection \cite{DBLP:conf/stoc/Charikar02} to determine joinability among pairs of attributes \cite{DBLP:conf/icde/BogatuFP020}. 
Hence, a pair of attributes will be ranked higher if the estimated overlapping of their instance sets is also high. Note, however, that such \textit{syntactic} definition does not discern pairs of attributes from different domains (e.g., there exist several musical bands that use city or state names), leading to a large number of false positives. While it might be feasible to manually discern such false positives on a handful of datasets, this task is unattainable at scale. 
In order to showcase the detrimental impact of using such measures to determine joinability, we designed an experiment collecting 138 datasets from open repositories such as Kaggle and OpenML\footnote{Repository available at \url{https://mydisk.cs.upc.edu/s/GeYwdYH7xsGqbaX}}. 
Precisely, we devised an heterogeneous collection of datasets ranging different topics, which yielded a total of 110,378 candidate pairs of textual attributes, where 4,404 of those have a containment higher or equal than $0.1$. We, then, manually labeled such pairs as either \textit{semantic} or \textit{syntactic}, distinguishing whether a pair of attributes share common values and, respectively, do or do not refer to the same concept in a shared domain. Such ground truth is publicly available to the community and available in the paper's companion website.

\begin{figure}
	\centering
	\begin{minipage}{0.5\linewidth}
		\centering
		\includegraphics[width=1\linewidth]{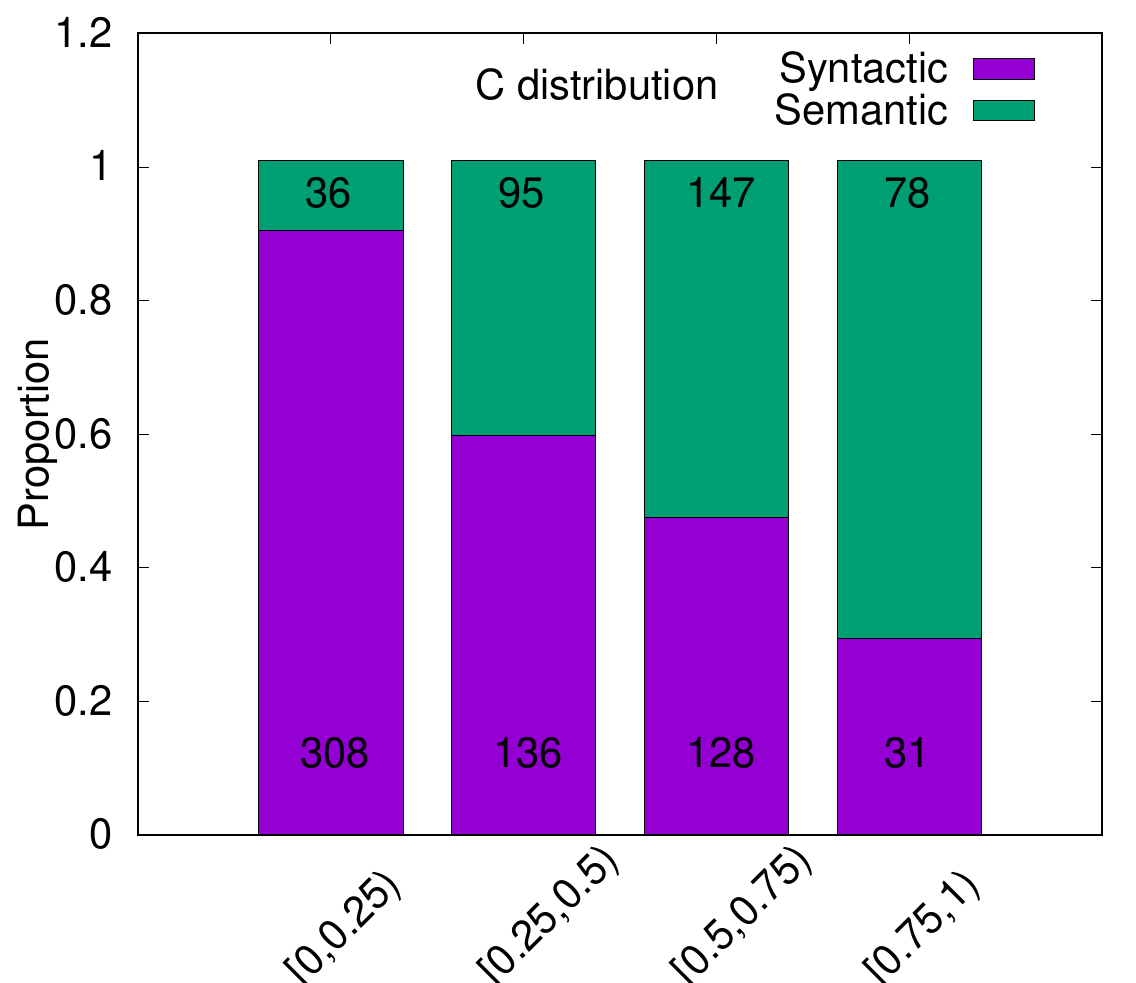} 
	\end{minipage}\hfill
	\begin{minipage}{0.5\linewidth}
		\centering
		\includegraphics[width=1\linewidth]{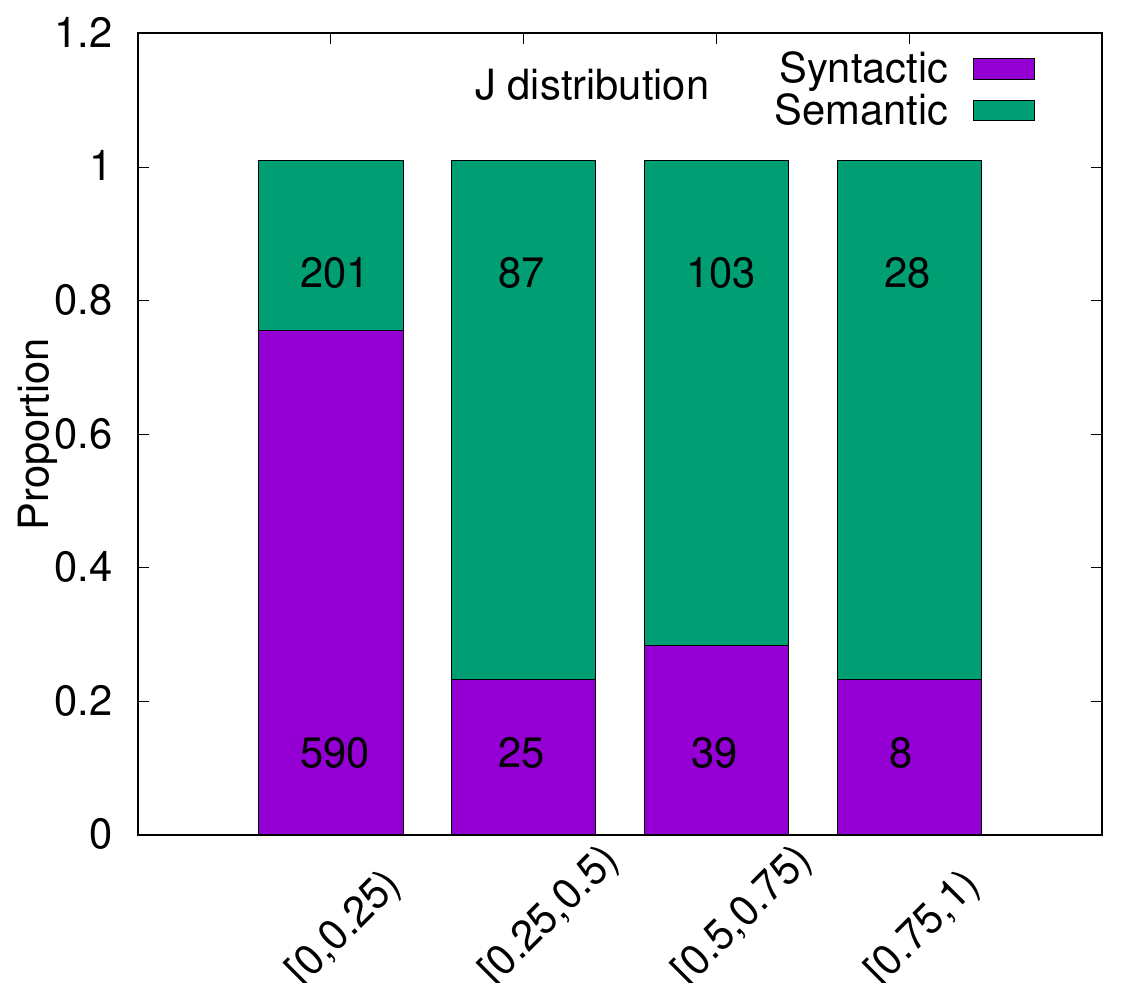} 
	\end{minipage}
	\caption{Proportion of syntactic and semantic pairs for different ranges of containment (left) and Jaccard (right) values. Labels in each bar denote the count of pairs in the range.}\label{fig:c_j_distribution}
\end{figure}

As shown in Figure \ref{fig:c_j_distribution}, even for high values of $C$ and $J$, the number of \textit{syntactic} pairs, and thus false positives, represent a substantial part. Then, in Table \ref{tab:c_vs_j}, we show the performance metrics (i.e., precision, recall and F-score) when considering different threshold values over $C$ and $J$ to discern syntactic (i.e., below the threshold) and semantic pairs (i.e., above the threshold) on the ground truth. We can observe that, overall, $C$ has a lower precision than $J$, indicating that it has a higher false positive rate. In contrast, $J$ has a lower recall than $C$, indicating that it has a higher false negative rate. Yet, overall, we can observe that in terms of F-score, which denotes the accuracy of the classifier, both metrics clearly fail at the task of identifying semantic pairs of joinable attributes.

\begin{table}[]
	\begin{tabular}{|ccc|ccc|}
		\hline
		\multicolumn{3}{|c|}{$\bm{C > 0.5}$}                                        & \multicolumn{3}{c|}{\textbf{$\bm{J > 0.5}$}}                                        \\ \hline
		\multicolumn{1}{|c|}{$P = 0.59$} & \multicolumn{1}{c|}{$R = 0.72$} & $F = 0.65$ & \multicolumn{1}{c|}{$P = 0.74$} & \multicolumn{1}{c|}{$R = 0.39$} & $F = 0.51$ \\ \hline
		\multicolumn{3}{|c|}{$\bm{C > 0.6}$}                                                 & \multicolumn{3}{c|}{$\bm{J > 0.6}$}                                                 \\ \hline
		\multicolumn{1}{|c|}{$P = 0.67$} & \multicolumn{1}{c|}{$R = 0.56$} & $F = 0.61$ & \multicolumn{1}{c|}{$P = 0.79$} & \multicolumn{1}{c|}{$R = 0.29$} & $F = 0.43$ \\ \hline
		\multicolumn{3}{|c|}{$\bm{C > 0.7}$}                                                 & \multicolumn{3}{c|}{$\bm{J > 0.7}$}                                                 \\ \hline
		\multicolumn{1}{|c|}{$P = 0.64$} & \multicolumn{1}{c|}{$R = 0.44$} & $F = 0.52$ & \multicolumn{1}{c|}{$P=0.78$}   & \multicolumn{1}{c|}{$R=0.20$}   & $F=0.32$   \\ \hline
		\multicolumn{3}{|c|}{$\bm{C > 0.8}$}                                                 & \multicolumn{3}{c|}{$\bm{J > 0.8}$}                                                 \\ \hline
		\multicolumn{1}{|c|}{$P=0.63$}   & \multicolumn{1}{c|}{$R=0.38$}   & $F=0.47$   & \multicolumn{1}{c|}{$P=0.75$}   & \multicolumn{1}{c|}{$R=0.17$}   & $F=0.28$   \\ \hline
		\multicolumn{3}{|c|}{$\bm{C > 0.9}$}                                                 & \multicolumn{3}{c|}{$\bm{J > 0.9}$}                                                 \\ \hline
		\multicolumn{1}{|c|}{$P=0.61$}   & \multicolumn{1}{c|}{$R=0.30$}   & $F=0.40$   & \multicolumn{1}{c|}{$P=0.80$}   & \multicolumn{1}{c|}{$R=0.16$}   & $F=0.26$   \\ \hline
		\multicolumn{3}{|c|}{$\bm{C\text{ }=\text{ }1.0}$}                                                   & \multicolumn{3}{c|}{$\bm{J\text{ }=\text{ }1.0}$}                                                   \\ \hline
		\multicolumn{1}{|c|}{$P=0.60$}   & \multicolumn{1}{c|}{$R=0.25$}   & $F=0.35$   & \multicolumn{1}{c|}{$P=0.75$}   & \multicolumn{1}{c|}{$R=0.11$}   & $F=0.20$   \\ \hline
	\end{tabular}
	\captionof{table}{Performance metrics of using different thresholds over $C$ and $J$ to identify semantic pairs. $P$, $R$ and $F$ denote, respectively, precision, recall and F-score} \label{tab:c_vs_j}
\end{table}

\subsection{Computing and accurately predicting the quality of a join}

The discussion above highlights the limitations of current data discovery approaches over large-scale scenarios. Indeed, the first challenge lies in the definition of a similarity measure such that it prioritizes pairs of attributes with large overlapping and shared domains, as an indicator of semantic relationship. Next, the second challenge is that of efficiently computing such measure at scale. As previously discussed, value and hash-based data discovery approaches do not scale well. Alternatively, comparison by profile methods are a better suit since they rely on the detection of similarities or discrepancies between profiles. Working with summaries instead of data values is much more efficient from a complexity point of view. Yet, despite the clear performance benefits of profile-based approaches, there is nowadays a large gap in the trade-off regarding the quality of their results mainly due to the adoption of rather basic profiles (e.g. \cite{DBLP:conf/icde/FernandezAKYMS18}) that do not accurately describe the underlying data or representative profiles (e.g. \cite{DBLP:journals/debu/ChenGHTD18}) that are used to discover a binary class (e.g. joinable or non-joinable). 
In order to overcome these issues, in this paper, we extend our previous vision paper \cite{DBLP:conf/edbt/0002N021a} and propose a novel approach to data discovery which aims to cover the gap generated by the low predictive performance of profile-based methods, as well as the limited precision and scalability of hash-based systems on large data lakes. 

We, first, propose a novel metric to measure the quality of a join. Opposite to the state-of-the-art, mostly focused on containment or Jaccard distance, we also consider the cardinality proportion between attributes as an indicator of a higher join quality. This allows us to get rid of a substantial amount of false positives, reducing the number of pairs to analyze. This is specially relevant in large-scale settings, where as shown in Figure \ref{fig:c_j_distribution}, the number of candidate pairs is too large to manually disregard false positives. 
Second, we propose a novel learning-based method based on profiles to discover 
joinable attributes for large-scale data lakes. Our assumptions apply to scenarios where data is denormalized and file formats embed tabular data (i.e., not nested). We rely on state-of-the-art relational data profiling techniques \cite{DBLP:journals/vldb/AbedjanGN15} to compute informative profiles for datasets. This task, which can be done offline and parallelized over distributed computing frameworks (e.g., Apache Spark), allows us to extract and model the underlying characteristics of attributes. Next, profiles are compared in order to predict their expected join quality. 
We show that our method is generalizable and that proposes a meaningful ranking of pairs of attributes based on the predicted join quality. We further show that our model is generalizable for data lake-alike settings.

\medskip

\noindent\textbf{Contributions.} We summarize our contributions as follows:
\begin{itemize}
	\item We introduce a quantitative metric for join quality, which considers containment and cardinality proportion between attributes.
	\item We learn a highly accurate and general model to predict and efficiently rank candidate pairs of joinable attributes.
	\item We extensively evaluate our approach to show it is scalable and outperforms the current state-of-the-art, yielding higher predictive performance results.
\end{itemize}

\noindent\textbf{Outline.}
The rest of the paper is structured as follows. We discuss related work and introduce the formal background, respectively in Sections \ref{sec:relatedwork} and \ref{sec:background}. Next, in Section \ref{sec:computingQ} we present the definition of the proposed quality metric, while Section \ref{sec:predictingQ} shows our approach to predicting it. Section \ref{sec:experiments} presents exhaustive experiments to showcase the effectiveness and scalability of our approach. We finally conclude our paper and present future work in Section \ref{sec:conclusions}. 

\section{Related Work}\label{sec:relatedwork}

In this section, we survey related work for each category identified in Table \ref{tab:relatedWork}.

\medskip

\noindent\textbf{Comparison by value.}
SilkMoth \cite{DBLP:journals/pvldb/DengKMS17} proposes a method to generate signatures from a subset of attribute tokens. To select an optimal subset, it uses an heuristic. Such signatures are used in an inverted index to prune the search space. Then, a verification step is required on the remaining candidates to discard those that do not hold for a certain similarity measure. This approach supports edit distance and Jaccard coefficient as similarity measures. It assumes all signatures fit in memory. JOSIE \cite{DBLP:conf/sigmod/ZhuDNM19} proposes to optimize the number of comparisons by scanning only the required values. Tokens are extracted from each attribute to create a dictionary and an inverted index. A ranked list is built from the $k$ most relevant candidate tables with highest containment, where attributes ranked at the top will have a larger number of common values. PPJoin \cite{DBLP:journals/tods/XiaoWLYW11} performs a different optimization by using prefix filtering to avoid computing similarity values for all possible values. This reduces the number of comparisons and hence improve efficiency. However, an inverted index requires a large space in memory. This approach proposes a similarity measure which combines tokens and characters.

\medskip

\noindent\textbf{Comparison by hash.} 
MinHash \cite{DBLP:conf/sequences/Broder97} uses the minwise hash function from the LSH collection, with a collision probability equal to the Jaccard similarity. This requires, for every value, to compute the MinHash signature $K$ times, where $K$'s magnitude is in the hundreds. This approach has a major limitation on performance, as well as a bias towards small sets introduced by the Jaccard similarity. To overcome this, under the observation that MinHash can be optimized providing a Jaccard threshold, LSH Ensemble \cite{DBLP:journals/pvldb/ZhuNPM16} proposes to use containment similarity and convert it to a Jaccard threshold. It focuses on finding attributes with a high containment similarity, that is, to cover as many values as possible. For efficient indexing, LSH Ensemble partitions the sets according to the set size. 
GB-KMV \cite{DBLP:conf/icde/YangZZH19} aims to reduce the number of false positives generated by LSH Ensemble. Further, it considers that additional information (e.g., attribute cardinalities and value frequencies) to offer better performance in estimating the containment similarity. Another approach that aims to tackle the drawbacks of MinHash is Lazo \cite{DBLP:conf/icde/FernandezMNM19}. Here, the Jaccard similarity is redefined to consider set cardinalities, which allows to estimate the containment similarity. Instead of computing $K$ times a hash function, Lazo implements the One Permutation Hashing (OPH) technique, hashing data values only once.  
A distinguishable feature of Lazo is that rehashing the entire dataset collection is not required when a new one is introduced. 
Aurum \cite{DBLP:conf/icde/FernandezAKYMS18}, represents relations between datasets and their attributes in a graph data structure (i.e., the \textit{Enterprise Knowledge Graph}). In this graph, attribute nodes are related if their hashing signatures, generated from their instances, are similar in an LSH index. To determine similarity it uses two metrics: Jaccard (i.e., MinHash similarity) and cosine (i.e., TF-IDF). Finally, the approach presented by D3L \cite{DBLP:conf/icde/BogatuFP020}, also employs LSH indexes generated from four kind of features (i.e., attribute names, values, formats and domain distributions) as well as word embeddings generated from the values. Hence, attribute joinability is based on the composition of the similarity of each of such five features.

\medskip

\noindent\textbf{Comparison by profile.}
LSD \cite{DBLP:journals/ml/DoanDH03} proposes a multi-strategy learning approach to automatically find related attributes among XML files. It applies multiple learner modules, where each module exploits different kind of information, either from schema or data values. Such predictions are combined to weigh each learner. LSD also exploits domain integrity constraints and user feedback. 
FlexMatcher \cite{DBLP:journals/debu/ChenGHTD18} extends LSD with more data types. A relevant aspect is that it considers pattern classifiers to filter data values. 
A limitation is that every time a discovery process is to be performed it requires to train new models providing a training sample of attributes that might join with the specific attribute. 
A different approach is SIMUBC \cite{DBLP:conf/esws/KejriwalM15a}, which aims to detect pairs of attributes sharing common values. SIMUBC extracts 28 kinds of metadata from attributes such as tokens, phonetic values or representatives. Such metadata are used to train Random Forest and Multilayer Perceptron models to predict whether two attributes are join candidates. To improve performance, weights are assigned to each model to compute the final prediction. A limitation of this work is that it requires to train the models each time a data discovery process is started. 
Then, PEXESO \cite{DBLP:conf/icde/DongT0O21} presents an approach to create high dimensional vectors (i.e., embeddings) from each record of a column. Then, attributes can be efficiently compared to each other via such embeddings. A major limitation of PEXESO is that it requires indexing in memory the embeddings of the complete data lake. To alleviate this issue, the paper presents partitioning techniques. 
The approach presented by DLN \cite{DBLP:journals/pvldb/BharadwajGBG21} is that of building a ML model to find join relationships from Microsoft's data lake Cosmos. The paper argues that two metadata-based features suffice to build a high quality model. On the one hand, the first feature is an embedding-enhanced column name similarity, for which they use word embeddings trained on software domain data and calculate the cosine similarity. On the other hand, the second feature is the column-name uniqueness, where the maximum ITF (inverse term frequency) of all individual tokens is used. Unfortunately, the paper does not provide reproducibility information or ground truth to compare with. 
Finally, WarpGate \cite{DBLP:journals/corr/abs-2212-14155}, is a prototype system that targets data discovery over cloud data warehouses by applying an embedding approach. These are built from columns with the objective that joinable columns will be closer in the higher dimensional embedding space. One of the limitations of WarpGate is, however, the runtime complexity of building such embedding spaces for large datasets.

\section{Preliminaries}\label{sec:background}

Here, we introduce the formal background of our approach.

\subsection{Measuring the quality of a join}

In this subsection, we fix the data model and formalize metrics for join quality.

\medskip

\noindent\textbf{Data repositories and datasets.}
A data repository $\mathcal{D}$ is a finite nonempty set of dataset names $\{ D_1, \ldots, D_m \}$, where each $D_i$ has a fixed arity $n_i$. Let $A$ be a set of attribute names, then each $D_i \in \mathcal{D}$ is associated to a tuple of attributes denoted by $att(D_i)$. Henceforth, we will assume that $\forall i,j : i \neq j  \rightarrow att(D_i) \cap att(D_j) = \emptyset$ (i.e., relations do not share attribute names), which if required can be done prefixing attribute names with their relation name. Then, we use $att(\mathcal{D})$ to refer to the set $\{ atts(D_1) \cup \ldots \cup atts(D_n) \}$. Then, let $V$ be a set of values, a tuple $t$ in $D_i$ is a function $t: att(D_i) \rightarrow V$. For any dataset $D_i$, $tuples(D_i)$ denotes the set of all tuples of $D_i$.

\medskip

\noindent\textbf{Joinable pairs.} Given two distinct datasets $D_a, D_b$ and a pair of attributes $\langle a,b \rangle$, such that $a \in att(D_a)$ and $b \in att(D_b)$ and value-sets $A$ and $B$, we say the pair $\langle a,b \rangle$ is \textit{syntactically joinable} if $A \cap B \neq \emptyset$. Following the definition from \cite{DBLP:conf/icde/KoutrasSIPBFLBK21}, we also say that such pair of attributes is \textit{semantically joinable} if they are \textit{syntactically joinable} and there exists a function $h: A \rightarrow B$ denoting semantic equivalence between attributes (i.e., both refer to the same concept). In practice, attributes with a semantic relationship also have a syntactic one. When this is not satisfied, as happens for the pair \textit{Country} (in Table~\ref{tab:happinessExample}) and \textit{Nation} (in Table~\ref{tab:expectancyExample}), we refer to this relationship as \textit{semantic non-syntactic}.

\medskip

\noindent\textbf{Quantifiable measures for joinability.} A quantifiable way to define that a pair of attributes are joinable is by using set-based coefficients (i.e., coefficients over well-defined collections of distinct values). As earlier discussed, two of the most commonly used coefficients are the inclusion coefficient ($C(A,B)$) and Jaccard coefficient ($J(A,B)$), which are formalized as:

\medskip

\begin{minipage}{13em}
	\begin{equation*} \label{eq:containmentScore}
		C(A,B) = \frac{ |A \cap B|}{|A|}
	\end{equation*}
\end{minipage}
\begin{minipage}{0em}
	\begin{equation*} \label{eq:jaccardSimilarity}
		J(A,B) = \frac{ |A \cap B|}{|A \cup B|}
	\end{equation*}
\end{minipage}

\medskip

Note Jaccard similarity is symmetric, thus it can be biased towards smaller sets. Oppositely, containment measures the relative size of the intersection of two sets over the size of one. Hence, such measure is asymmetric.

\medskip

\noindent\textbf{Join quality metric.} A join quality metric is a function $\mathcal{Q}: (\mathcal{A},\mathcal{B}) \rightarrow \mathbb{R}$ from the set of all sets of values $\mathcal{A}$ and $\mathcal{B}$ to the set of real numbers, such that, for any set of values $A,B,C,D$ it holds that $\mathcal{Q}(A,B) > \mathcal{Q}(C,D)$ if the pair $\langle A,B \rangle$ is semantically joinable and the pair $\langle C,D \rangle$ is syntactically joinable. 

Note this generalization allows to include containment and Jaccard as join quality functions, yet it does not consider the possibility to rank joinable pairs of the same kind. This is due to the fact that there is no consensus in the literature on which metric is best. Hence, one of the contributions of this paper is on the proposal of a novel metric to determine the quality of a join.

\subsection{Predicting the quality of a join}\label{sec:statement}

Since the computation of a join quality measure $\mathcal{Q}$ might be unattainable at scale, we also consider the join discovery problem as a predictive task.

\medskip 

\noindent\textbf{Profiles.}
A unary profile $P_u$ for an attribute $A$, referred as $P_u(A)$ is a set of meta-features $\{ m_1, \ldots, m_n \}$. Each $m_i$ is a summary or statistic about the structure or content of $A$ (e.g., number of distinct values). We also consider binary profiles, which are meta-features that denote characteristics of a relationship between pairs of attributes. Hence, we define a binary profile $P_b$ for a pair of attributes $A,B$, denoted $P_b(A,B)$, as a set of meta-features (e.g., Levenshtein distance between attribute names).

\medskip

\noindent\textbf{Join quality prediction function.} A join quality prediction function is a function $\mathcal{P}: (\overline{P_u},\overline{P_u}^\prime,\overline{P_b}) \rightarrow \mathbb{R}$ from a triple defined by the set of all unary profiles, from both the reference and candidate attributes, and the set of all binary profiles, to the set of real numbers, such that, for any set of values $A,B,C,D$ if $\mathcal{Q}(A,B) > \mathcal{Q}(C,D)$ then $\mathcal{P}(P_u(A),P_u(B),P_b(A,B)) > \mathcal{P}(P_u(C),P_u(D),P_b(C,D))$. 

\medskip

\noindent\textbf{Problem statement.} We now formalize the predictive join discovery problem. The goal is to discover a ranking (i.e., a partially-ordered set) of equi-join predicates based on their predicted join quality.

\begin{definition}[Discovery-by-attribute]
	Let $A_q$ be a query attribute, $D_{ref}$ a reference dataset where $A_q \in att(D_{ref})$, and $\mathcal{D}$ a data repository where $D_{ref} \notin \mathcal{D}$; obtain a partially-ordered set of joinable pairs $R$ of the form $R=\{ \langle A_q, A_1 \rangle, \ldots, \langle A_q, A_n \rangle \}$, where $A_1, \ldots, A_n \in att(\mathcal{D})$ such that $\forall \langle A_q, A_i \rangle, \langle A_q, A_j \rangle \in R: \langle A_q, A_i \rangle \succ \langle A_q, A_j \rangle \implies \mathcal{P}(P_u(A_q), P_u(A_i), P_b(A_q,A_i))$ $\geq \mathcal{P}(P_u(A_q), P_u(A_j),$ $P_b(A_q,A_j))$.
\end{definition}

The remainder of the paper is devoted to \textit{a)} present a novel instantiation of join quality metric (see Section \ref{sec:computingQ}), and \textit{b)} present an approach to instantiate the join quality prediction function (see Section \ref{sec:predictingQ}).

\section{A novel metric to measure the quality of a join}\label{sec:computingQ}

Here, we define a novel metric to determine the quality of a join.

\subsection{On the cardinality proportion's role}

\begin{table*}[]
	\begin{center}
		\begin{subtable}{.275\linewidth}
			\centering
			\caption{\label{tab:tourismIncome} $D_{ref}$ -- Tourism income in Spain}
			\resizebox{\textwidth}{!}{
				\begin{tabular}{|c|c|c|}
					\hline
					\textbf{City} & \textbf{Seaside} & \textbf{Amount}  \\
					\hline
					Barcelona & Y & 350M \\
					\hline
					Girona & Y & 110M \\
					\hline
					Lleida & N & 75M \\
					\hline
					Tarragona & Y & 83M \\
					\hline
					\ldots & \ldots & \ldots \\
					\hline
				\end{tabular}
			}
		\end{subtable}
		\begin{subtable}{.01\linewidth}
		\end{subtable}
		\begin{subtable}{.4\linewidth}
			\caption{\label{tab:demographicEU} $D_1$ -- EU demographic data}
			\centering
			\resizebox{\textwidth}{!}{
				\begin{tabular}{|c|c|c|c|}
					\hline
					\textbf{Unit} & \textbf{Population} & \textbf{Avg. salary} & \textbf{Cost of living} \\
					\hline
					Antwerp & 1,120,000 & 44,000 \euro & 2,896 \euro \\
					\hline
					Barcelona & 1,620,343 & 31,000 \euro & 2,422 \euro \\
					\hline
					Berlin & 4,725,000 & 49,000 \euro & 2,737 \euro  \\
					\hline
					Bristol & 1,157,937 & 30,000 \pounds & 2,397 \pounds \\
					\hline
					\ldots & \ldots & \ldots & \ldots \\
					\hline
				\end{tabular}
			}
		\end{subtable}%
		\begin{subtable}{.01\linewidth}
			\hspace*{\fill}
		\end{subtable}%
		\begin{subtable}{.3\linewidth}
			\centering
			\caption{\label{tab:demographicWorld} $D_2$ -- Worldwide demographic data}
			\resizebox{\textwidth}{!}{
				\begin{tabular}{|c|c|c|}
					\hline
					\textbf{Name} & \textbf{Country} & \textbf{ Population } \\
					\hline
					Barcelona & Spain & 1,620,343 \\
					\hline
					Canberra & Australia & 426,704 \\
					\hline
					Chicago & United States & 2,695,598 \\
					\hline
					Curitiba & Brasil & 1,908,359 \\
					\hline
					\ldots & \ldots & \ldots \\
					\hline
				\end{tabular}
			}
		\end{subtable}
	\end{center}
	\caption{ \label{tab:motivateJoinQuality} A reference dataset ($D_{ref}$) and two candidate datasets to be joined. $D_1$ is curated with extensive data at European level, while $D_2$ is curated at the worldwide level with less details}
\end{table*}

Unlike the state-of-the-art, which mainly uses containment and Jaccard similarities to decide the degree of joinability among pairs of attributes, we aim to define a metric to measure the expected join quality. 
As shown in Table \ref{tab:c_vs_j}, containment yields better results to determine the joinability of a pair attributes with respect to Jaccard. Yet, we make the observation that datasets on a data lake do not relate to each other as in a relational database. In such scenarios, it is common to find datasets with few data values in common. In order to exemplify this idea, let us consider the datasets depicted in Table \ref{tab:motivateJoinQuality}. In this example, the reference dataset $D_{ref}$ might be joined with any of the two candidate datasets $D_1$ (at the EU level) and $D_2$ (worldwide). Current approaches would propose both as positive pairs, since they yield the same containment. However, we aim at distinguishing the join quality between them and use their \emph{cardinality proportion} for that purpose, which is defined by the following expression:

\begin{equation*}
	K(A,B) = \dfrac{ min(|A|,|B|)}{max(|A|,|B|)}
\end{equation*}

\medskip

Let us, then consider the following cardinalities corresponding to the city attributes (which are the only relevant ones to join): |$City$| = 8,124, |$Unit$| = 20,000 and |$Name$| = 54,500, respectively belonging to $D_{ref}$, $D_1$ and $D_2$. We use the cardinality proportion as a measure to infer whether their data granularities are similar. In this sense, the joinable attribute in $D_2$ is much larger than that in |$D_{ref}$| and yields a worse proportion compared to $D_1$, and thus should be ranked lower. Importantly, we assume these datasets store independently generated events and such big difference in their cardinality indicates they embed different semantics or sit at different granularity levels. In general, such situations are a source of false positives for current solutions, specially, when considering small tables.

\subsection{Definition of an empirical metric}

\begin{figure}[htbp]
	\begin{center}
		\includegraphics[width=.8\linewidth]{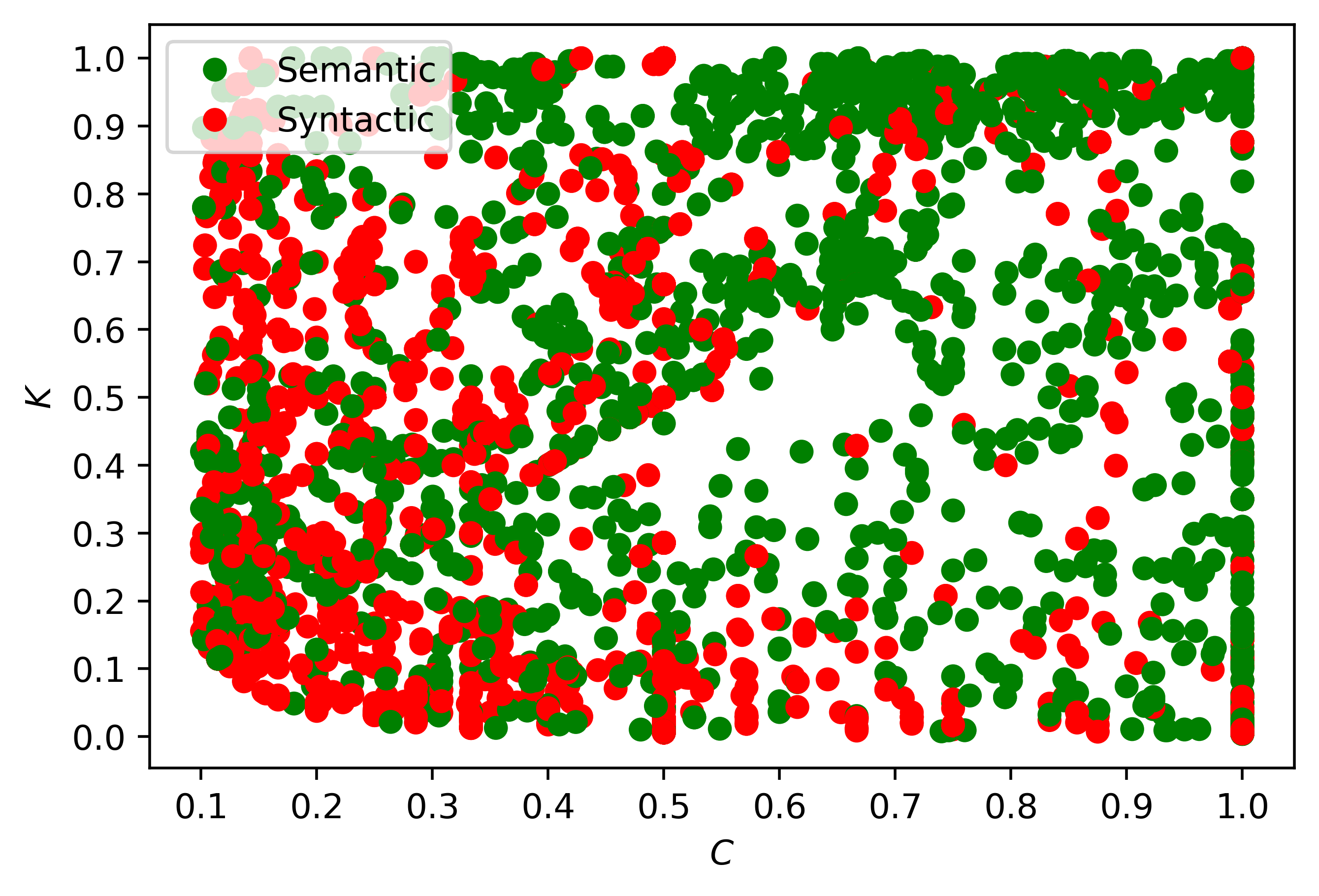}
		\caption{Distribution of syntactically and semantically-joinable pairs in the ground truth over $C$ and $K$}
		\label{fig:C_K}
	\end{center}
\end{figure}

We, then, follow an empirical approach to define the join quality metric. This is, from a set of quantifications drafted from a sample, we aim to derive a measure that can generalize to a population \cite{wilks1961some}. In our setting, from the manually-labeled ground truth used to conduct the experiment depicted in Figure \ref{fig:c_j_distribution}, we observe how the containment and cardinality proportion values relate for both syntactically and semantically-joinable pairs. Indeed, as shown in Figure \ref{fig:C_K}, the rationale that the cardinality proportion is a valid metric to discern false positives provided by the containment holds. As observed, most of the syntactically-joinable pairs have a value of $C < 0.5$, yet for those that are above such threshold most of them lie below the range $K < 0.5$. In other words, we can identify semantically-joinable pairs when both $C$ and $K$ are closer to 1.

From such observations in the ground truth, a naive approach to discern syntactically and semantically-joinable pairs would be that expressed by the following expression, which would yield $1$ if a pair is semantically-joinable and $0$ otherwise:

\begin{equation*}
	Q(A,B) = 
	\begin{cases}
		1, &\text{if $C(A,B) \geq \tfrac{1}{2}$ and $K(A,B) \geq \tfrac{1}{2}$}\\
		0, & \hfill\text{otherwise}
	\end{cases}
\end{equation*}

Yet this metric is still limited, as is the case for the other ones in the state-of-the-art, on its ability to rank pairs that are of the same kind. We, hence, generalize and propose a multi-class metric to determine the quality of a join based on multiple quality levels $L$ (i.e., degrees of joinability) as defined by the following expression:

\begin{equation*}
	Q(A,B,L) = max(i) \in [0, \ldots, L] | C(A,B) \geq (1-\dfrac{i}{L}) \wedge K(A,B) \geq \dfrac{1}{2^i}
\end{equation*}

\begin{figure}[h!]
	\centering
	\begin{minipage}{0.5\linewidth}
		\centering
		\includegraphics[width=1\linewidth]{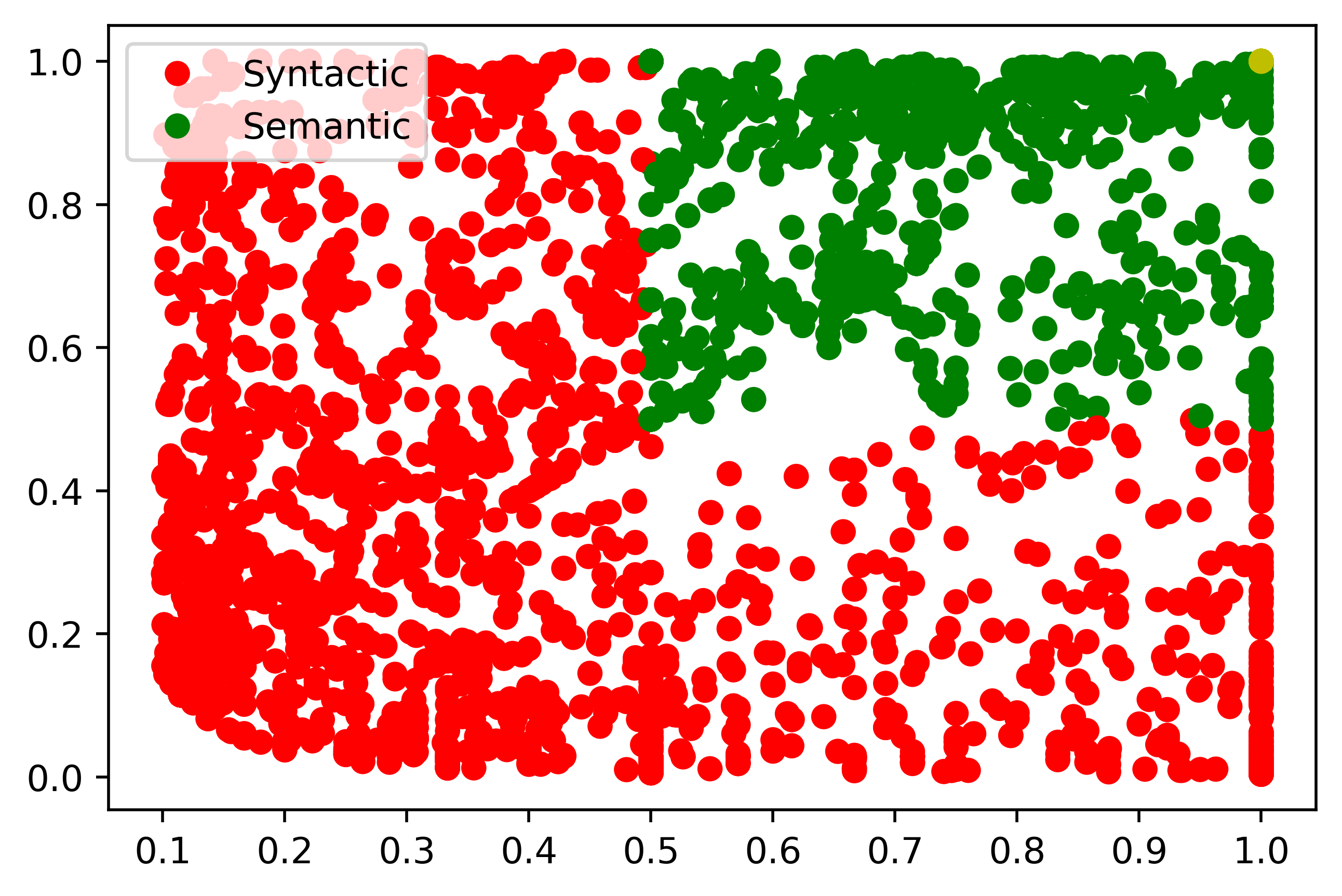} 
	\end{minipage}\hfill
	\begin{minipage}{0.5\linewidth}
		\centering
		\includegraphics[width=1\linewidth]{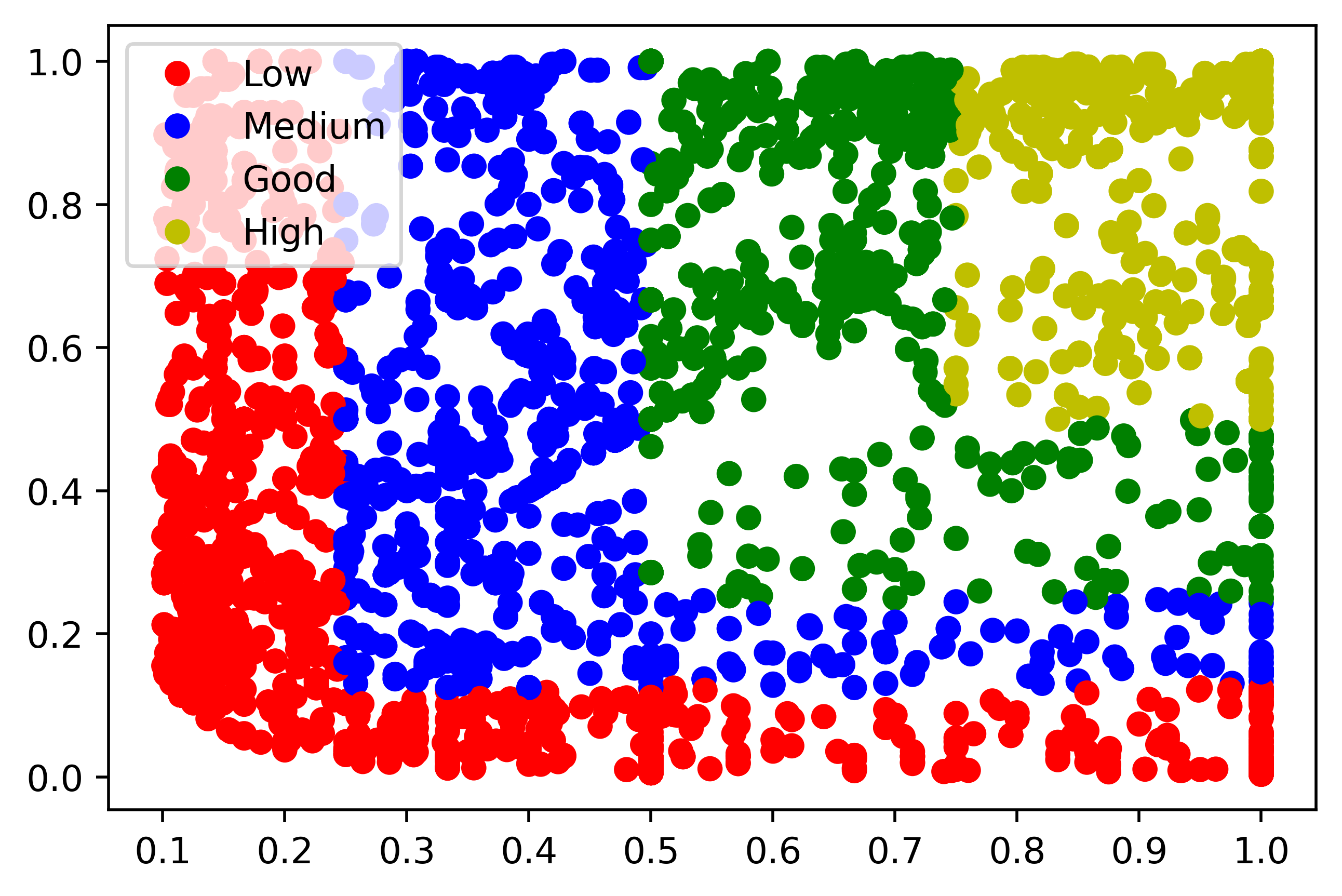} 
	\end{minipage}
	\caption{Areas identified by the quality metric for $L=2$ (left) and $L=4$ (right)}\label{fig:quality_levels}
\end{figure}

The intuition of $Q(A,B,L)$ is that of defining equally-sized buckets for $C(A,B)$ and constraint them using $K(A,B)$. Figure \ref{fig:quality_levels}, depicts the areas defined by such quality metric for the case of $L=2$ (which is equivalent to $Q(A,B)$ earlier introduced) and $L=4$. The latter uses richer labels, for this case denoted as \textit{Low}, \textit{Medium}, \textit{Good}, and \textit{High} for the different levels of quality, respectively 0.25, 0.5, 0.75 and 1 (note that we ignore the value 0 in the chart). Hence, a pair labeled \textit{High} will always be ranked higher than one labeled \textit{Good}. The interpretation of such metric is that of measuring the quality of the join's output from $A$'s perspective under equi-join semantics (i.e., under the semantics of left-semijoin conjunctive queries). This is, how the number of elements in $A$ will be reduced after doing a join operation using $B$.

Take again the example from Table \ref{tab:motivateJoinQuality} and consider the following containment values $C(City,Unit) = 0.8$ and $C(City,Name) = 0.95$, and cardinality proportion values $K(City,Unit) = 0.40$ and $K(City,Name) = 0.15$.  
Note that, although the containment is very high in both cases, the constraint on cardinality proportions allows to rank in a higher position the first, denoting a more interesting join result.

To showcase the benefit of considering the cardinality proportion to complement the containment, consider the following extreme case, which is not uncommon on large-scale automated data discovery scenarios. Consider the two datasets depicted in Table \ref{tab:extreme}, the former ($D_s$) listing the opening hours of stores and the latter ($D_m$) movies and their directors. Let us assume |$Store$| = 3 and |$Movie$| is above a million movies. Solutions exclusively considering containment would qualify the pair $\langle D_s.Store, D_m.Movie \rangle$ pair as a high quality join, given the 2/3 containment (which would be even higher if we consider approximate joins). 
Yet, this is clearly a false positive. 
Considering the cardinality proportion, our quality metric would penalize its ranking and assign a low value to this candidate pair. Note that the Jaccard index is able to deal with this case, yet as shown in Table \ref{tab:c_vs_j} it generally has a high false negative rate deeming it suboptimal.

	\begin{table}[h!]
		\begin{center}
			\begin{subtable}{.48\linewidth}
				\centering
				\resizebox{\textwidth}{!}{
					\begin{tabular}{|c|c|c|}
						\hline
						\textbf{Store} & \textbf{Open} & \textbf{Close} \\
						\hline
						Chicago & 8am & 18pm  \\
						\hline
						Casablanca & 9:30am & 20pm  \\
						\hline
						Paris & 9am & 18pm  \\
						\hline
					\end{tabular}
				}
			\end{subtable}
			\begin{subtable}{.01\linewidth}
			\end{subtable}
			\begin{subtable}{.48\linewidth}
				
				\centering
				\resizebox{\textwidth}{!}{
					\begin{tabular}{|c|c|}
						\hline
						\textbf{Movie} & \textbf{Director} \\
						\hline
						An American in Paris & G. Gershwin  \\
						\hline
						Casablanca & M. Curtiz  \\
						\hline
						Chicago & R. Marshall \\
						\hline
						\ldots & \ldots \\
						\hline
					\end{tabular}
				}
			\end{subtable}%
			\begin{subtable}{.01\linewidth}
				\hspace*{\fill}
			\end{subtable}%
		\end{center}
		\caption{ \label{tab:extreme} $D_{s}$, store schedules (left), and  $D_{m}$, movies and their directors (right)}
	\end{table}
	
\subsection{A continuous metric for join quality}
	
Despite the ability of $Q(A,B,L)$ to assign quality levels beyond binary ones, the output of such metric is discrete, and thus the rankings it generates are bounded by $L$. In order to overcome this issue, we aim to provide a generalization of such discrete metric into a continuous one $Q(A,B)$ in the continuous range $[0,1]$. The approach we follow is that of plotting the empirical distribution function (\textit{edf}) of $Q(A,B,L)$ for some value of $L$, and then fit a continuous probability distribution on it. Empirical distributions are functions that describe a sample of observations for a given variable, while probability distributions are functions that yield the probability of different possible values for a variable. We distinguish between probability density functions (\textit{pdf}), which yield the probability that a random variable takes on a certain value, and cumulative distribution functions (\textit{cdf}) which yield the probability that a random variable takes on a value less than or equal to a certain value. Thus, we are precisely interested in the latter. The challenge is to determine what distribution function better fits our metric.
	
\medskip
	
\noindent\textbf{Fitting a Gaussian distribution.} The most notorious kind of probability distribution is the Gaussian distribution, also known as the normal distribution, $\mathcal{N}(\mu,\sigma^{2})$. The \textit{pdf} of the normal distribution is defined by the following expression:
	
\begin{equation*} \label{eq:pdf_N}
	\displaystyle{pdf(x;\mu,\sigma^{2})=\frac{1}{\sqrt{2\pi\sigma^{2}}}e^{^{\frac{-(x-\mu )^{2}}{2\sigma ^{2}}}}}
\end{equation*}
	
\medskip 
	
The \textit{cdf} of the normal distribution is defined as:
	
\begin{equation*} \label{eq:cdf_N}
	\displaystyle{cdf(x;\mu,\sigma^{2})=\frac{1}{\sqrt{2\pi}}\int_{-\infty }^{\frac{x-\mu}{\sigma}}e^{\frac{-t^{2}}{2}}dt}
\end{equation*}
	
\medskip
	
Yet, since we are working with a two-dimensional function for $C$ and $K$ we must consider the case for the multivariate normal distribution. Assuming $C$ and $K$ are independent, the \textit{cdf} truncated in the range $[a,b]$ for a two-dimensional Gaussian for $C$ and $K$ (i.e., $cdf_{\mathrm{CK}} \left(c,k\right)$) is equal to the product of the individual \textit{cdf}s of $C$ and $K$ (i.e., $cdf_C \left(c\right)cdf_K \left(k\right)$), which is defined as:
	
\begin{equation*}
	cdf_{\mathrm{CK}} \left(c,k\right)=\frac{\Phi \left(\frac{c-\mu_c }{\sigma_c }\right)-\Phi \left(\frac{a-\mu_c }{\sigma_c }\right)}{\Phi \left(\frac{b-\mu_c }{\sigma_c }\right)-\Phi \left(\frac{a-\mu_c }{\sigma_c }\right)}\frac{\Phi \left(\frac{k-\mu_k }{\sigma_k }\right)-\Phi \left(\frac{a-\mu_k }{\sigma_k }\right)}{\Phi \left(\frac{b-\mu_k }{\sigma_k }\right)-\Phi \left(\frac{a-\mu_k }{\sigma_k }\right)}
\end{equation*}
	
where $\Phi(x)$ is the univariate \textit{cdf} of the normal distribution defined by the following expression:
	
\begin{equation*}
	\Phi \left(x\right)=\frac{1}{2}\left(1+\mathrm{erf}\left(\frac{x}{2}\right)\right)=\frac{1}{2}\left(1+\frac{2}{\sqrt{\pi }}\int_0^{\frac{x}{2}} e^{-t^2 } \mathrm{dt}\right)
\end{equation*}
	
\medskip

Hence, the challenge reduces to finding the mean values $\mu_c$ and $\mu_k$, which determine the offset of the distribution, and the covariance matrix $\Sigma = \big(\begin{smallmatrix}
		\sigma_c & 0\\
		0 & \sigma_k
	\end{smallmatrix}\big)$, which determine the shape of the distribution, such that it best fits the discrete quality function $Q(A,B,L)$. To do so, we consider the Wasserstein metric which is a distance function defined over probability distributions. Hence, the goal is to find the optimal values such that minimize the Wasserstein distance. Such task, which can be evaluated in a brute-force manner over a discrete range of values, yielded the following values: $\mu_c = 0$, $\mu_k = 0.44$, and $\Sigma = \begin{pmatrix}
		0.19 & 0\\ 
		0 & 0.28
	\end{pmatrix}$.
	
	\medskip
	
\begin{figure}[h!]
		\centering
		\begin{minipage}{0.5\linewidth}
			\centering
			\includegraphics[width=1\linewidth]{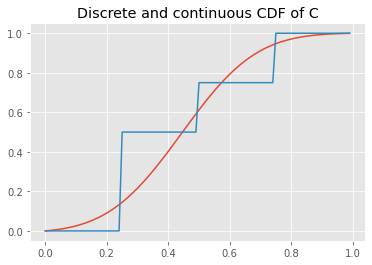} 
		\end{minipage}\hfill
		\begin{minipage}{0.5\linewidth}
			\centering
			\includegraphics[width=1\linewidth]{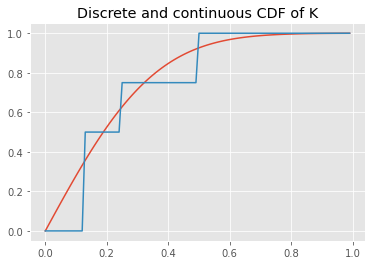} 
		\end{minipage}
		\caption{Resulting \textit{cdf}s that minimize the Wasserstein distance over the \textit{edf} of $C$ (left) and $K$ (right) for $Q(A,B,4)$}\label{fig:cdfs}
	\end{figure}

	Figure \ref{fig:cdfs} depicts resulting fit of the \textit{cdf} of the normal distribution over $C$ and over $K$ using the previously introduced values, which is superposed over the \textit{edf} of $Q(A,B,4)$.
	
	\medskip
	
	\noindent\textbf{Continuous quality metric with strictness levels}. Since the previously presented values for mean and covariance matrix have been derived from our ground truth, we finally consider the possibility to include a certain degree of variability on the resulting quality metric. To that end, we consider the \textit{strictness} score $s$ and consider three possible values: \textit{relaxed} (i.e., $s=0$), \textit{balanced} (i.e., $s=0.25$) and \textit{strict} (i.e., $s=0.5$). Such score will vary the value of $\mu_c$ (i.e., the mean of the containment value), which, as can be observed in Figure \ref{fig:quality_levels}, is the dominating factor in our metric. Hence, the resulting continuous join quality metric $Q(A,B,s)$ is defined as:
	
	\begin{equation*}
		cdf \Big( \mu_C+s,\Sigma[0][0],0,1,C(A,B) \Big) cdf \Big( \mu_K,\Sigma[1][1],0,1,K(A,B) \Big)
	\end{equation*}
	
\begin{figure*}[h!]
		\begin{center}
			\includegraphics[width=1\linewidth]{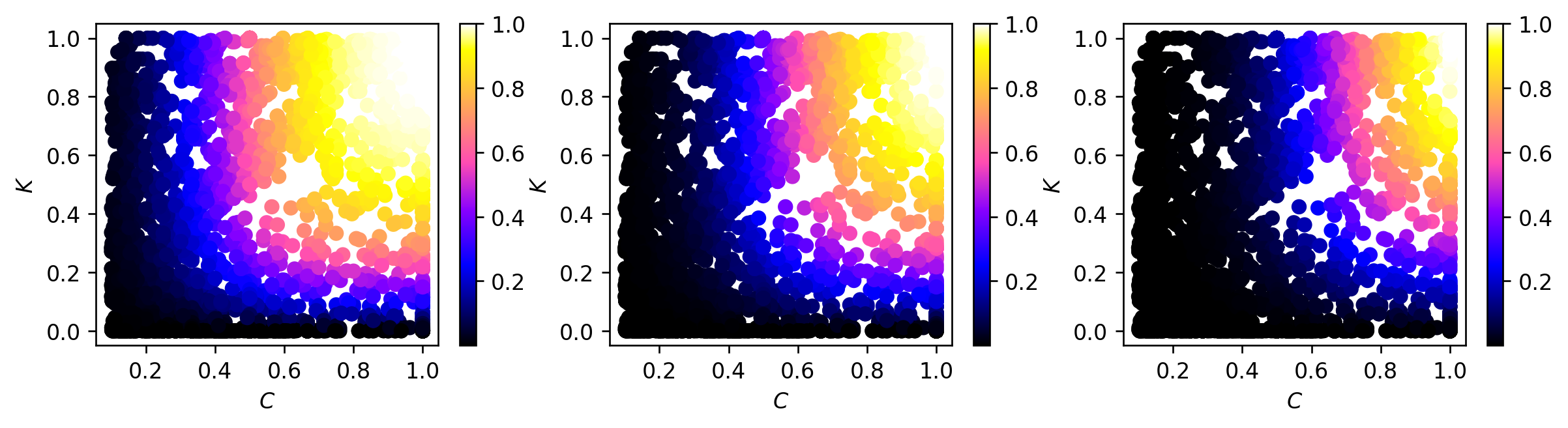}
			\caption{Continuous quality in the ground truth for $Q(A,B,0)$, $Q(A,B,0.25)$, and $Q(A,B,0.5)$}
			\label{fig:continuousQualityInGroundTruth}
		\end{center}
	\end{figure*}

	Figure \ref{fig:continuousQualityInGroundTruth}, shows the application of such metric for the different strictness values considered over our ground truth.
	
\section{Predicting the quality of a join} \label{sec:predictingQ}

In this section, we describe our approach to predict the join quality metric introduced in Section \ref{sec:computingQ}.

\subsection{Attribute profiling}

Profiles are composed of meta-features that represent the underlying characteristics of attributes. Such profiles are the key ingredient for high accuracy predictions, thus we require an exhaustive summary of attributes. Hence, we base our profiling on state-of-the-art relational data profiling techniques \cite{DBLP:journals/vldb/AbedjanGN15}. We distinguish meta-features corresponding to unary and binary profiles. We further distinguish the former into meta-features modeling cardinalities, value distribution and syntax. A summary of all the implemented meta-features is depicted in Table \ref{tab:metadataList}. Although for space reasons it has not been included here, we validated by means of a principal component analysis the relevance of all meta-features towards meaningful profiling of attributes.

\begin{table*}[]
	\centering
	\resizebox{\textwidth}{!}{
		\begin{tabular}{|c|c|l|c|}
			\hline
			\textbf{Category} & \textbf{Meta-feature} & \multicolumn{1}{c|}{\textbf{Description}} & \textbf{Norm.?} \\
			\hline
			\multirow{4}{*}{Cardinalities}                                                  & Cardinality              & Number of distinct values within an attribute & Yes                                                                                                \\
			\cline{2-4}
			& Uniqueness & Measures if the attribute contains unique values   & No \\
			\cline{2-4}
			& Incompleteness           & Measures the number of missing values                                                                                                      & No                                                                                                 \\
			\cline{2-4}
			& Entropy                  & Measures the variety of an attribute                                                                                                  & Yes                                                                                                \\
			\hline
			\multirow{11}{*}{\begin{tabular}[c]{@{}c@{}}Value \\distribution \end{tabular}} & Average frequency        & The average value of the frequency distribution count                                                                                      & Yes                                                                                                \\
			\cline{2-4}
			& Min frequency            & The minimum value of the frequency distribution count                                                                                      & Yes                                                                                                \\
			\cline{2-4}
			& Max frequency            & The maximum value of the frequency distribution count                                                                                      & Yes                                                                                                \\
			\cline{2-4}
			& SD frequency             & The standard deviation of the frequency distribution count                                                                                 & Yes                                                                                                \\
			\cline{2-4}
			& Octiles                  & The octiles (quantiles) of the frequency distribution in percentages                                                                       & No                                                                                                 \\
			\cline{2-4}
			& Min perc frequency       & The minimum value of the frequency distribution in percentages                                                                             & No                                                                                                 \\
			\cline{2-4}
			& Max perc frequency       & The maximum value of the frequency distribution in percentages                                                                             & No                                                                                                 \\
			\cline{2-4}
			& SD perc frequency        & The standard deviation of the frequency distribution in percentages                                                                        & No                                                                                                 \\
			\cline{2-4}
			& Constancy                & Frequency of the most frequent value divided by number of rows                                                                             & No                                                                                                 \\
			\cline{2-4}
			& Frequent words           & The 10 most frequent words                                                                                                                 & No                                                                                                 \\
			\cline{2-4}
			& Soundex                  & The 10 most frequent words in soundex representation                                                                                       & No                                                                                                 \\
			\hline
			\multirow{12}{*}{Syntactic}                                                      & Data type                & \begin{tabular}[c]{@{}l@{}}The data type of the attribute (i.e., numeric, alphanumeric, alphabetic, \\nonAlphanumeric, or datetime) \end{tabular}    & No                                                                                                 \\
			\cline{2-4}
			& Specific type            & \begin{tabular}[c]{@{}l@{}}The specific type of the attribute (i.e., phone, email, url, ip, username, or phrases) \end{tabular}                    & No                                                                                                 \\
			\cline{2-4}
			& Percentage data type     & The percentage for each data type detected in the data values                                                                              & No                                                                                                 \\
			\cline{2-4}
			& Percentage specific type & The percentage for each specific type detected in the data values                                                                          & No                                                                                                 \\
			\cline{2-4}
			& Longest string           & The number of characters in the longest string                                                                                             & Yes                                                                                                \\
			\cline{2-4}
			& Shortest string          & The number of characters in the shortest value in the attribute                                                                            & Yes                                                                                                \\
			\cline{2-4}
			& Average string           & Average length of the strings in term of characters                                                                                        & Yes                                                                                                \\
			\cline{2-4}
			& Number words             & The number of words in the attribute                                                                                                       & Yes                                                                                                \\
			\cline{2-4}
			& Average words            & The average words in the attribute                                                                                                         & Yes                                                                                                \\
			\cline{2-4}
			& Min words                & The minimum words in the attribute                                                                                                         & Yes                                                                                                \\
			\cline{2-4}
			& Max words                & The maximum words in the attribute                                                                                                         & Yes                                                                                                \\
			\cline{2-4}
			& SD words                 & The standard deviation in the attribute                                                                                                    & Yes                                                                                                \\
			\hline
			\multirow{3}{*}{\begin{tabular}[c]{@{}c@{}}Pair\\metadata \end{tabular}}      & Best containment          & The containment score assuming all distinct values are covered                                                                           & No                                                                                                 \\
			\cline{2-4}
			& Flipped containment       & \begin{tabular}[c]{@{}l@{}}Containment assuming all distinct values are covered divided by max cardinality \end{tabular} & No                                                                                                 \\
			\cline{2-4}
			& Name distance  & Measures the difference of two attribute names using Levenshtein distance & No                                                                                                 \\
			\hline
		\end{tabular}
	}
	\captionof{table}{Meta-features composing a profile} \label{tab:metadataList}
\end{table*}

\medskip

\noindent\textbf{Cardinalities.} These provide a broad view of an attribute. 
Uniqueness, which is computed dividing the number of distinct values by the cardinality, allows us to quantify the extent of duplicated values. A uniqueness smaller than $1$ indicates there exists duplicate values, hence we can identify which attributes have high redundancies. We can also detect incompleteness, which is determined by the number of missing values divided by the cardinality. This produces a value in the range $[0,1]$, where values closer to $1$ denote the attribute has a high percentage of missing values. Finally, entropy, also referred as \textit{diversity index}, measures the variety of data in an attribute. 

\medskip

\noindent\textbf{Value distribution.}
Here, we exploit information in a fine-grained manner by using a frequency distribution of the attribute values, either by count or percentage. Despite its simplicity, the frequency distribution of column values exposes insightful characteristics, such as how often a value occurs. We compute frequency metrics (e.g., in the form of octiles), and descriptive statistics (e.g., mean, standard deviation, etc.) to characterize the distribution of the data. We also take a sample of the ten most frequent values.

\medskip

\noindent\textbf{Syntax.}
This category of unary metadata describes the patterns of data. 
These meta-features include information regarding the length of values in characters, such as the length of the longest and shortest value, and the average length. We also compute information regarding syntactic consistency, such as format and data type. This aids to give meaning to the attribute's content. We also infer the data type of an attribute, in a broad and fine-grained manner. Broad data types are generic descriptions such as numeric, alphabetic, alphanumeric, dateTime, non-alphanumeric, etc. However, we also extract its fine-grained type to extract what content is the attribute representing. To this end, we use regular expressions that allow us to model usernames, 
phrases, 
phones, etc. In order to improve the quality of meta-features in this category, we preprocess values to lowercase, remove accents and special symbols. 

\medskip

\noindent\textbf{Binary meta-features.}
We also extract meta-features regarding pairs of attributes. 
We use Levenshtein distance to obtain the similarity between pairs of attribute names \cite{1966SPhD...10..707L}. This is normalized by the length of the largest string.

\subsection{Comparing profiles}

Before comparing profiles and due to the fact attribute meta-features are represented in different magnitudes, we normalize them to guarantee a meaningful comparison. As shown in Table \ref{tab:metadataList}, we consider a large amount of meta-features that require normalization. 
Two common normalization techniques are Min-Max and Z-score. The former consists on rescaling data into the range $[0,1]$, This technique, however, is sensitive to outliers which will lay on the boundaries. Oppositely, Z-score normalization overcomes this issue by rescaling values to have a mean of $0$ and a standard deviation of $1$. For this reason, we use Z-score to normalize meta-features. The following equation depicts the normalization process, which requires the mean and standard deviation of the metadata, which is computed from all the values of each attribute to be compared.

\mathchardef\mhyphen="2D
\begin{equation*} \label{eq:z_score}
	Z\mhyphen score = \frac{(x - \mu )}{\sigma }\
\end{equation*}

After normalizing each meta-feature we compute the distances among pairs of attributes. Here, we also compute binary meta-features. The result of this stage is a set of distance vectors $D$ where, for each $D_i$, values closer to $0$ denote high similarities.

\medskip

\noindent\textbf{Training a regression model.} Once the distance vectors are computed, we can train the predictive model. Precisely, the goal is to train a model so that, for a pair of attributes $A,B$, its prediction (i.e.,  $\mathcal{P}(P_u(A),P_u(B),P_b(A,B))$) is highly correlated to the true join quality (i.e., $Q(A,B,s)$). For that, we fixed the intermediate value of $s=0.25$, and evaluated several regression models performing a hyperparameter grid search to minimize the error. Precisely, we built models based on: linear regression, ensemble learning, support vector regression, and multilayer perceptrons (MLP). The resulting best model, in terms of highest coefficient of determination $R^2$, was an MLP with ReLU function, a single hidden layer with dimension 100 and a value of $\alpha = 0.0001$. This model provided a $R^2 = 0.8831$.

\section{Experimental evaluation}\label{sec:experiments}

In this section, we present the evaluation of our approach. On the one hand, we evaluate and compare the ability of the model to generalize and discover quality joins with respect to state-of-the-art solutions. We, also, evaluate and compare its scalability. In order to present transparent experiments and guarantee the reproducibility of results, we created an informative companion website\footnote{{\url{https://www.essi.upc.edu/dtim/nextiajd/}}}. There, it is possible to find all necessary resources (i.e., source code, datasets, and detailed instructions) needed to reproduce the presented results.

We have implemented the profiling phase of \name~as an extension of Apache Spark.  
The runtime methods  
are implemented as new operators over the structured data processing library SparkSQL. We leverage on the Catalyst optimizer to efficiently compute the profiles and compare them. 
The predictive model is available as a \textit{Pickle}, which allows to easily adapt it in other projects. 

\subsection{Generalizability of our approach}\label{sec:Generalizability}

Since we empirically derive a join quality metric from ground truth, the first question is whether it is applicable to other data lake settings. Thus, the objective of this first experiment is to show the generizability of our approach. To that end, we perform experiments on datasets independently-generated from those selected in our ground truth and assess our metric's performance.

\medskip

\noindent\textbf{Methodology.} We consider the GitTables dataset \cite{DBLP:journals/corr/abs-2106-07258} as data repository to evaluate our approach. GitTables is a large-scale corpus of 1M relational tables extracted from CSV files in GitHub repositories. Each table is provided as a Parquet file, which comes with metadata associated. On average tables have 25 columns and 209 rows. Of special interest for us are the \textit{semantic annotation types} each attribute has, which tell the probability a column is similar to a semantic type in a knowledge base (e.g., DBpedia). As described in \cite{DBLP:journals/corr/abs-2106-07258}, these are annotated using a pretrained FastText model, and the annotation corresponds to the most similar semantic type. We, then, constructed a ground truth (i.e., calculated our join quality metric with a strictness level $s=0.25$) considering those attributes with a semantic similarity equal to $1.0$ that have the same semantic type (e.g., clusters of \textit{author}, \textit{city}, \textit{name}, etc.). Since evaluating all possible combinations of attributes over the complete corpus of GitTables is unattainable (i.e., it is quadratic in the total number of attributes), we reduced the search space to the \texttt{abstraction\_tables} and \texttt{dwarf\_tables} datasets, which represent the ones with highest volume.

\medskip

\noindent\textbf{Results.} We evaluated our predictive model on the annotated GitTables ground truth, which yielded a mean squared error (MSE) value of 0.04, and a mean absolute error (MAE) value of 0.13. Since the predictive model was trained from an independently-generated ground truth, we consider these error scores to be highly acceptable. Then, we also evaluated the ability of the predictive model to discern syntactically and semantically-joinable pairs following the same approach as in Table \ref{tab:c_vs_j}. Hence, Table \ref{tab:gittables} depicts the predictive perfomance metrics of using different thresholds to determine semantically-joinable pairs over the GitTables dataset. From these results, we can conclude that the precision of our metric is monotonically increasing with higher threshold values, while maintaining a constant high recall.

\begin{table}[h!]
	\begin{tabular}{|cccc|}
		\hline
		\multicolumn{4}{|c|}{$\bm{Q > 0.5}$}  \\ \hline
		\multicolumn{1}{|c|}{$P = 0.40$} & \multicolumn{1}{c|}{$R = 0.87$} & \multicolumn{1}{c|}{$F = 0.55$} & $A = 0.97$ \\ \hline
		\multicolumn{4}{|c|}{$\bm{Q > 0.6}$}  \\ \hline		
		\multicolumn{1}{|c|}{$P = 0.49$} & \multicolumn{1}{c|}{$R = 0.85$} & \multicolumn{1}{c|}{$F = 0.62$} & $A = 0.98$ \\ \hline
		\multicolumn{4}{|c|}{$\bm{Q > 0.7}$}  \\ \hline
		\multicolumn{1}{|c|}{$P = 0.58$} & \multicolumn{1}{c|}{$R = 0.85$} & \multicolumn{1}{c|}{$F = 0.69$} & $A = 0.98$ \\ \hline
		\multicolumn{4}{|c|}{$\bm{Q > 0.8}$}  \\ \hline
		\multicolumn{1}{|c|}{$P = 0.68$} & \multicolumn{1}{c|}{$R = 0.87$} & \multicolumn{1}{c|}{$F = 0.76$} & $A = 0.98$ \\ \hline
		\multicolumn{4}{|c|}{$\bm{Q > 0.9}$}  \\ \hline
		\multicolumn{1}{|c|}{$P = 0.76$} & \multicolumn{1}{c|}{$R = 0.88$} & \multicolumn{1}{c|}{$F = 0.82$} & $A = 0.99$ \\ \hline								
	\end{tabular}
	\captionof{table}{Performance metrics of using different thresholds over $Q$ to predict semantically-joinable pairs. $P$, $R$, $F$, and $A$ denote, respectively, precision, recall, F-score, and accuracy} \label{tab:gittables}
\end{table}

\begin{figure*}[t!]
	\begin{center}
		\includegraphics[width=1\linewidth]{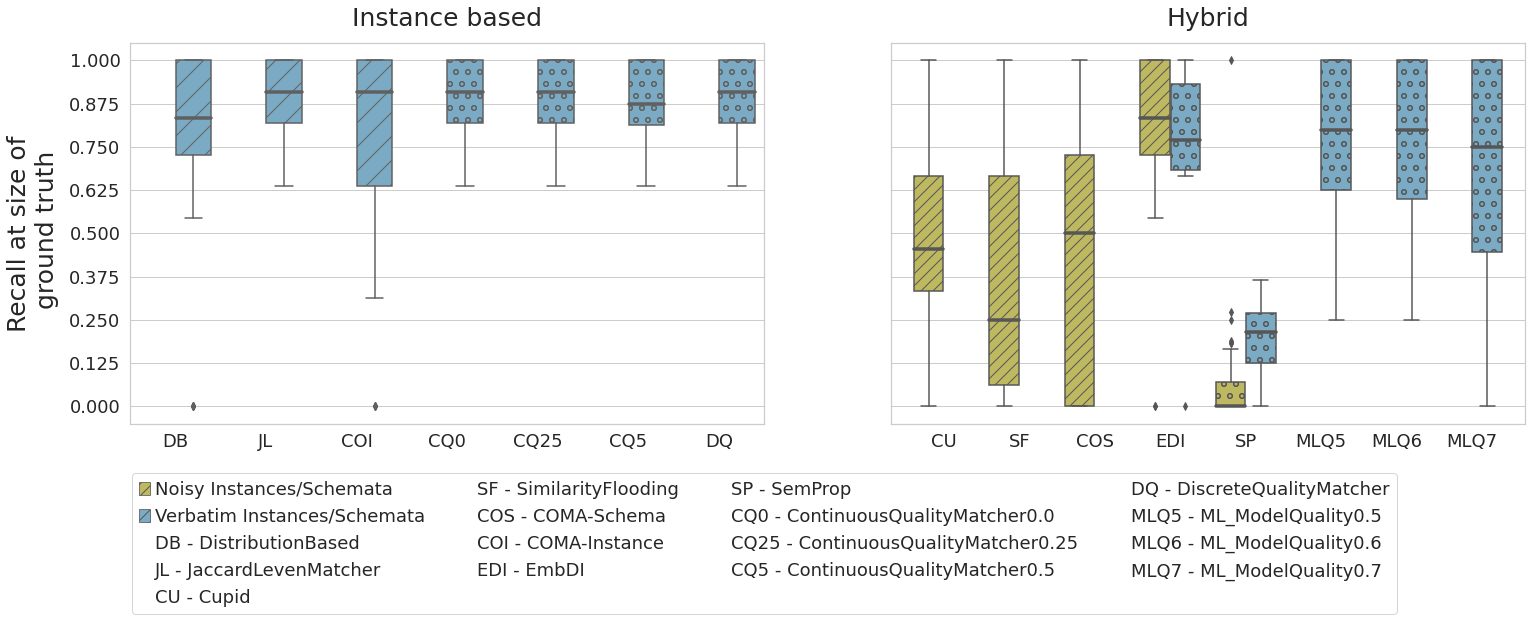}
		\caption{Recall at size of ground truth scores provided by Valentine}
		\label{fig:valentine_recallSizeGroundTruth}
	\end{center}
\end{figure*}
\subsection{Comparison with the state-of-the-art (schema matching)}

Here, we experimentally compare our quality metric and its associated predictive model. 

\medskip

\noindent\textbf{Methodology.} In this experiment we rely on the Valentine experimental suite for data discovery \cite{DBLP:conf/icde/KoutrasSIPBFLBK21}. Valentine provides the implementation of seven schema matching algorithms (which are base for data discovery methods), together with a collection of ground truth annotated datasets. The datasets consider datasets manually annotated as well as automatically annotated. We, precisely, focus on their \textit{joinable} scenario in Valentine, which is equivalent to our definition of semantically-joinable pairs. We extended Valentine to incorporate new matchers for \textit{a)} our discrete quality metric with $L=4$; \textit{b)} our continuous quality metric with a strictness level of $s=0.25$; and \textit{c)} the learned predictive model of \textit{b)}. Since Valentine's ground truth is labeled on a binary class (i.e., joinable or non-joinable), we considered variants of the previously discussed matchers with different threshold values. We extended Valentine's plots, which are represented as boxplots, with our performance metrics under both instance-based and hybrid scenarios. Since our approach does not consider data transformations to calculate the join quality, we did not consider the scenario where Valentine incorporates noise in the instances and schema, and focused only on the verbatim setting.

\medskip

\noindent\textbf{Results.} We, first report on the \textit{recall at size of ground truth} evaluation metric. This is a metric that, if the size of ground truth is $N$, then only the first $N$ proposed matches will be considered for the recall calculation. This is a metric that allows to assess the quality of the ranking a method provides. Figure \ref{fig:valentine_recallSizeGroundTruth}, depicts the obtained results by means of boxplots. On the left-hand side, we encounter the instance-based methods, which are those that compute their rankings based on the data values of each attribute. Hence, here, we present the results for our discrete quality metric (i.e., \textit{DQ}), and the continuous one with three thresholds (i.e., \textit{CQ}). Here, we can observe that the quality of the ranking provided by all our matcher variants is equal or better than state-of-the-art instance-based methods. This is another indicator of how our proposed metric generalizes, since here it is evaluated on an independently labeled ground truth. Yet, more interestingly, we shift our attention to the right-hand side of Figure \ref{fig:valentine_recallSizeGroundTruth}, depicting the recall at size of ground truth for hybrid approaches (i.e., those that make use of schema and auxiliary data structures previously generated from the instances). Here, we compare the matchers encapsulating our predictive models (i.e., \textit{ML}) with 0.5, 0.6 and 0.7 as threshold to determine joinability. As it can be observed, our predictive approach yields better rankings than most competitors only being in pair with EmbDI \cite{DBLP:conf/sigmod/CappuzzoPT20}, a method that finds relationships between columns comparing their previously trained embeddings.

\begin{figure*}[h!]
	\begin{center}
		\includegraphics[width=1\linewidth]{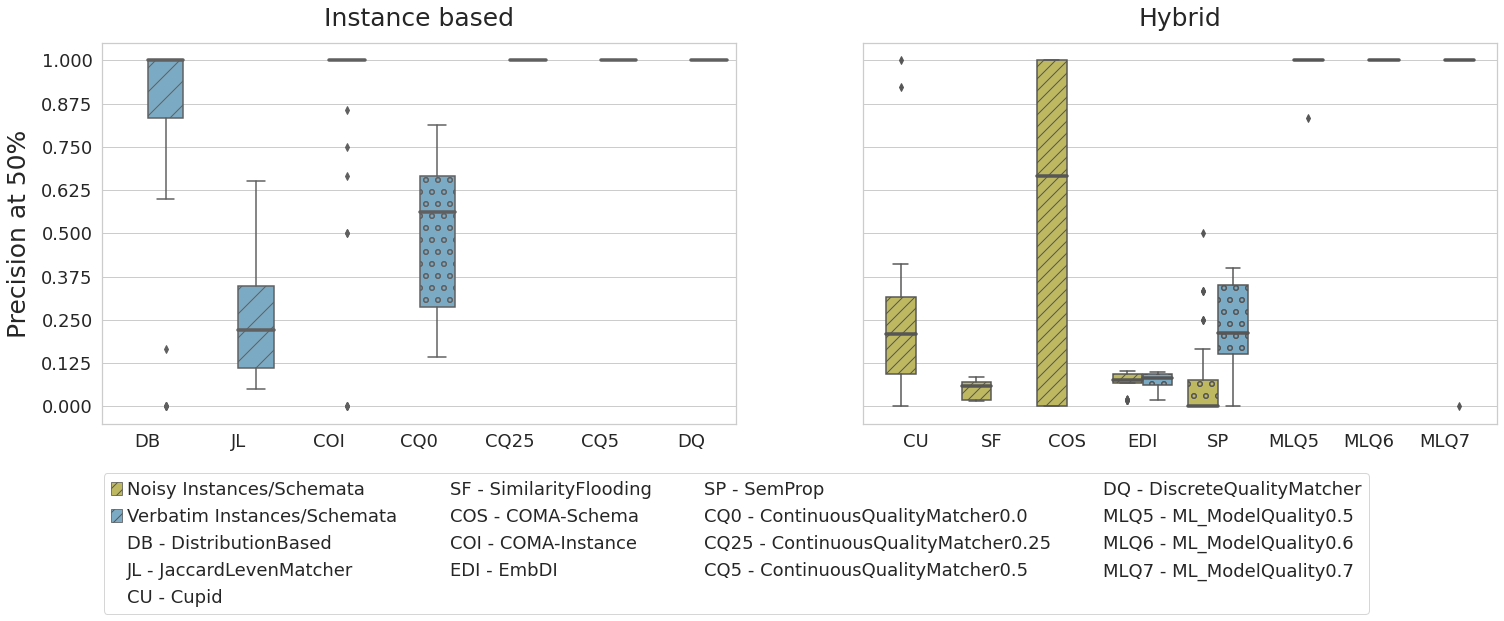}
		\caption{Precision at $50\%$ scores provided by Valentine}
		\label{fig:valentine_precision50}
	\end{center}
\end{figure*}

Next, we compare the efectiveness of the approaches by means of the \textit{precision at 50\%} performance metric. This is a metric that computes the precision only over the top 50\% of the provided ranking, which allows to determine the level of false positives a matcher yields. As shown in Figure \ref{fig:valentine_precision50}, both our calculated instance-based and predicted metrics clearly outperform the state-of-the-art in terms of precision results. This is, all pairs returned by our approach are indeed true positives. Surprisingly, approaches such as EmbDI which have high recall scores, clearly fail on effectively identifying true positive matches. Concerning our approach, the obtained precision results are extremely good, with only few outliers provided by the predicted metric. After examining Valentine's benchmark, we believe that this is due to the fact that its ground truth datasets have a lack of heterogeneity and are not representative of a data lake environment, as opposite to the GitTables scenario evaluated in Section \ref{sec:Generalizability}.

\subsection{Comparison with the state-of-the-art (data discovery)}\label{sec:comparisondatadiscovery}
Here, we provide a comparison with state-of-the-art data discovery systems which are not part of the Valentine suite. Precisely, we compare \name~with Aurum \cite{DBLP:conf/icde/FernandezAKYMS18}, D3L \cite{DBLP:conf/icde/BogatuFP020}, and WarpGate \cite{DBLP:journals/corr/abs-2212-14155}. These are systems whose course code is openly available. Unfortunately, no other solutions in the realm of approximate data discovery (i.e., those based on hash or profiles) could be considered due to the fact that \textit{a)} the code is openly available but it cannot be reproduced due to outdated dependencies or it is hardcoded for specific environments, or \textit{b)} they assume unrealistic data lake scenarios expecting all datasets to be loaded and pre-processed in a relational database.  

\medskip

\begin{figure}[]
	\begin{center}
		\includegraphics[width=1\linewidth]{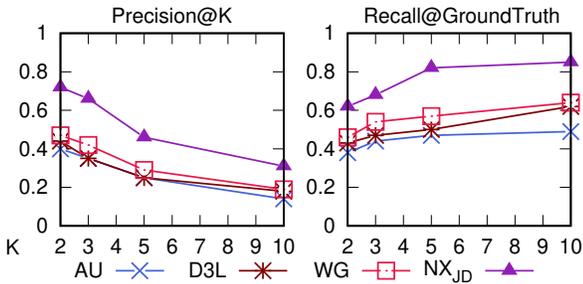}
		\vspace{-2em}
		\caption{Precision and recall scores on testbed $S$ for state-of-the art data discovery systems}
		\label{fig:S_results}
	\end{center}
\end{figure}

\noindent\textbf{Methodology.} For evaluation purposes, we collected 139 independent datasets from those used for the ground truth. We further divided such datasets into 4 testbeds: extra-small \textit{XS} ($0-1$ MB), small \textit{S} ($1 - 100$ MB), medium \textit{M} ($100$ MB $- 1$ GB) and large \textit{L} ($>1$ GB). Respectively, these contain 28, 46, 46, and 19 datasets, and 159, 590, 600, and 331 string attributes. For each testbed, manual ground truth was generated with a quality level according to the quality of the join (from 1 to 4), where those above 3 are considered semantically-joinable. Such testbeds are also available in the paper's companion website. The ability to rank joinable pairs on all the systems was assessed over top-K queries, where we report \textit{Precision@K} and \textit{Recall@GroundTruth}. The former scoring the ratio of true positives over K, and the latter scoring the ratio of true positives over the size of the ground truth for each query. To that end, for each candidate pair, we measure the quality metric with a \textit{balanced} strictness level.

\medskip

\begin{figure}[]
	\begin{center}
		\includegraphics[width=1\linewidth]{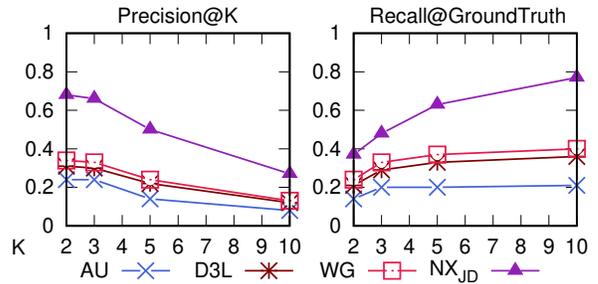}
		\vspace{-2em}
		\caption{Precision and recall scores on testbed $M$ for state-of-the art data discovery systems}
		\label{fig:M_results}
	\end{center}
\end{figure}

\noindent\textbf{Results.} We build on the openly available experimental results reported in \cite{DBLP:journals/corr/abs-2212-14155}, where Aurum, D3L and WarpGate are assessed over testsbeds \textit{S} and \textit{M}, since these are the largest ones in terms of datasets and ground truth defined in this paper. Then, in Figures \ref{fig:S_results} and \ref{fig:M_results} we report, respectively, top-K results for such two testbeds (note AU stands for Aurum, WG for WarpGate, and $\text{NX}_{\text{JD}}$ for \name). From the obtained results, we can see that \name~consistently outperforms the alternative systems in terms of precision and recall as the value of $k$ increases. Thus, \name~provides more accurate rankings than the alternatives with respect to the available ground truth. 
It is important, however, to remark that due to the lack of community consensus on how to quantify joinability for data discovery, the compared systems do not target the same objective metric. Thus, in this experiment we have followed the definition proposed in this paper, which is motivated to address large-scale data lake scenarios where data are highly denormalized and file formats embed tabular data. On such metric, \name~outperforms the alternative systems in terms of precision and recall for top-K queries.

\subsection{Scalability}
As previously discussed, our most intensive task with regard to computational resources is the generation of attribute profiles from datasets. Thus, here we perform stress tests of this component by means of two experiments.

\medskip

\noindent\textbf{Methodology.}
We generated a 10GB base CSV file with 5 columns and systematically extended it in batches of 10GBs, up to 60GBs. Next, we followed a similar strategy with regard to columns. We created a 20GB base file that was systematically extended with a new duplicate column each time. The resulting files were stored in a Hadoop HDFS cluster, using the default block size and replication parameters. In order to simulate a realistic large-scale scenario, we also converted each input file to Apache Parquet\footnote{\url{https://parquet.apache.org/}} format. Parquet is an specialized hybrid layout that fragments data into row group partitions (i.e., physically-independent columns), while it also embeds numerous statistics to optimize queries. To evaluate the scalability of our approach in terms of distribution, we compute the profiling runtime using $n$ Spark workers (cf. HDFS datanodes) in the range $1 \ldots 3$.

\medskip

\begin{figure}[t!]
	\begin{center}
		\includegraphics[width=1\linewidth]{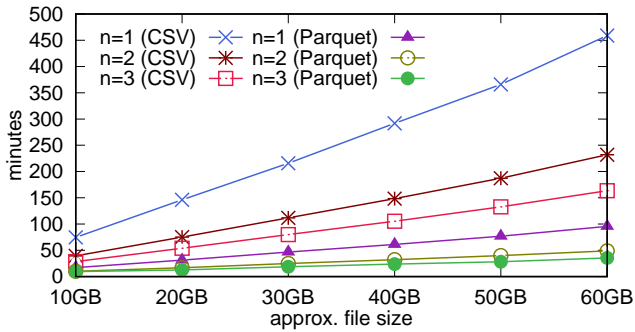}
		\vspace{-2em}
		\caption{Profiling runtime over an increasing file size}
		\label{fig:horizontalScalability}
	\end{center}
\end{figure}

\begin{figure}[t!]
	\begin{center}
		\includegraphics[width=1\linewidth]{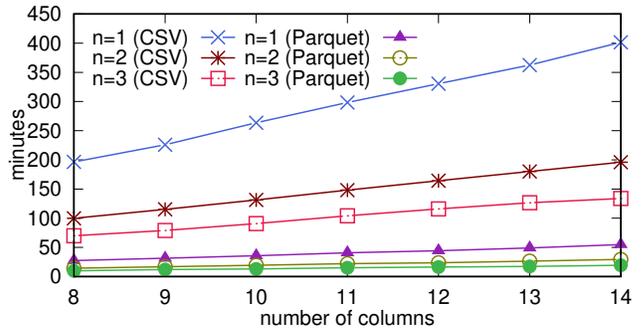}
		\vspace{-2em}
		\caption{Profiling runtime over an increasing number of columns}
		\label{fig:verticalScalability}
	\end{center}
\end{figure}

\noindent\textbf{Results.}
Figure \ref{fig:horizontalScalability} depicts the profiling runtime for an increasing file size. Regardless of the number of workers and data format used, the runtime linearly scales with the file size. As expected, profiling Parquet files are much more efficient than CSV ones (i.e., an approximate 4x to 5x speed-up), as we can benefit from statistics and compression when computing certain meta-features. 
As depicted in Figure~\ref{fig:verticalScalability}, we can also observe that the profiling runtime trend scales linearly with the number of columns. Similarly to the previous case, using Parquet significantly speeds up the process, here with a 7x to 8x factor. 

Finally, in Table \ref{tab:profileSize}, we show the average profile size per each testbed from those earlier presented. The disk usage is proportional to both the number of rows and columns. Although the number of columns is the leading factor for the profiling size, the dataset cardinality impacts on the size of some meta-features (e.g., frequent words, soundex, etc.). In any case, the profile sizes are reasonable. Thus, they can be precomputed offline and stored together with the dataset as metadata. The only exception to this would be binary meta-features. As final conclusion, these experiments show that our approach does not introduce any blocking factor hindering parallelism and can fully benefit from it.

\begin{table}[h!]
	\centering
	\resizebox{0.3\textwidth}{!}{
		\begin{tabular}{|c|c|}
			\hline
			\textbf{Testbed} & \textbf{Average profile size (KBs)} \\
			\hline
			$XS$ & $132$ \\
			\hline
			$S$ & $343$ \\
			\hline
			$M$ & $282$ \\
			\hline
			$L$ & $398.1$ \\
			\hline
		\end{tabular}
	}
	\caption{\label{tab:profileSize} Average profile size per testbed}
\end{table}

\section{Conclusions and future work}\label{sec:conclusions}

We have presented a novel learning-based approach for data discovery on large-scale repositories of heterogeneous, independently created datasets. Our work is motivated by (i) the poor predictive performance of current profile-based solutions, and (ii) the inability to scale-up of hash-based approaches, as well as their low precision, which is undesirable for large-scale scenarios. In order to overcome these limitations, we propose a scalable method yielding good precision, and grounded on a novel qualitative definition of join quality.  
We have experimentally shown that our approach outperforms the state-of-the-art data discovery approaches in terms of predictive and runtime performance. 
As future work, we look for adapting our approach to detect semantic non-syntactic join relationships (i.e., requiring some simple transformation on the values before joining). Based on such predictions, the system should be able to propose the required transformations to join.  

\balance

\begin{acks}
The authors are grateful to Tianji Cong for kindly providing the raw experimental results for the experiment reported in Section \ref{sec:comparisondatadiscovery}. This work was partly supported by the DOGO4ML project, funded by the Spanish Ministerio de Ciencia e Innovación under project PID2020-117191RB-I00 / AEI/10.13039/501100011033. Sergi Nadal is partly supported by the Spanish Ministerio de Ciencia e Innovación, as well as the European Union - NextGenerationEU, under project FJC2020-045809-I / AEI/10.13039/501100011033.
\end{acks}

\bibliographystyle{ACM-Reference-Format}
\bibliography{main}  

\end{document}